\documentclass[twocolumn,tighten]{aastex62}
\hypersetup{linkcolor=red,citecolor=blue,filecolor=cyan,urlcolor=magenta}
\usepackage{amsmath}
\usepackage{natbib}

\newcommand{\feh}{{\mathrm{[Fe/H]}}}
\newcommand{\gsix}{\object[G 64-12]{G~64-12}}
\newcommand{\hdeight}{\object[HD 84937]{HD~84937}}
\newcommand{\jtwo}{J1202$-$0020}
\newcommand{\jeight}{J1208$+$0024}

\newcommand{\jtwolong}{SDSS J120220.91$-$002038.9}
\newcommand{\jeightlong}{SDSS J120825.36$+$002440.4}

\newcommand{\degree}{$^{\circ}$}
\newcommand{\loggf}{\mbox{$\log gf$}}
\newcommand{\kmsec}{\mbox{km~s$^{\rm -1}$}}
\newcommand{\logg}{\mbox{log~{\it g}}}
\newcommand{\msun}{\mbox{$M_{\odot}$}}
\newcommand{\teff}{\mbox{$T_{\rm eff}$}}
\newcommand{\vt}{\mbox{$v_{\rm t}$}}
\newcommand{\rpro}{\mbox{{\it r}-process}}

\newcommand{\rv}{$V_{r}$}
\defcitealias{ibata19a}{I19}

\shorttitle{Spectroscopy of Sylgr Stream Stars}
\shortauthors{Roederer \& Gnedin}
\accepted{for publication in the Astrophysical Journal}

\begin{document}

\title{%
High Resolution Optical Spectroscopy of Stars 
in the Sylgr Stellar Stream\footnote{%
This paper includes data gathered with the 6.5~meter 
Magellan Telescopes located at Las Campanas Observatory, Chile.}
}

\author{Ian U.\ Roederer}
\affiliation{%
Department of Astronomy, University of Michigan,
1085 S.\ University Ave., Ann Arbor, MI 48109, USA}
\affiliation{%
Joint Institute for Nuclear Astrophysics -- Center for the
Evolution of the Elements (JINA-CEE), USA}
\email{Email:\ iur@umich.edu}

\author{Oleg Y.\ Gnedin}
\affiliation{%
Department of Astronomy, University of Michigan,
1085 S.\ University Ave., Ann Arbor, MI 48109, USA}

\begin{abstract}
We observe two metal-poor main sequence stars that are members
of the recently-discovered Sylgr stellar stream.
We present radial velocities, stellar parameters, and abundances
for 13 elements
derived from high-resolution optical spectra collected using the
Magellan Inamori Kyocera Echelle spectrograph.
The two stars have identical compositions
(within 0.13~dex or $1.2\sigma$) 
among all elements detected.
Both stars are very metal poor
([Fe/H]~$= -2.92 \pm 0.06$).
Neither star is highly enhanced in C
([C/Fe]~$< +1.0$).
Both stars are enhanced in the $\alpha$ elements Mg, Si, and Ca
([$\alpha$/Fe]~$= +0.32 \pm 0.06$), and
ratios among Na, Al, and all Fe-group elements are typical for other
stars in the halo and ultra-faint 
and dwarf spheroidal galaxies at this metallicity.
Sr is mildly enhanced
([Sr/Fe]~$= +0.22 \pm 0.11$),
but Ba is not enhanced
([Ba/Fe]~$< -0.4$),
indicating that these stars do not contain
high levels of neutron-capture elements.
The Li abundances match those found in metal-poor 
unevolved field stars and globular clusters
($\log\epsilon$(Li)~$= 2.05 \pm 0.07$), 
which implies that environment
is not a dominant factor in determining the Li content of
metal-poor stars.
The chemical compositions of these two stars
cannot distinguish whether the progenitor of the Sylgr stream
was a dwarf galaxy or a globular cluster.
If the progenitor was a dwarf galaxy,
the stream may originate from 
a dense region such as a nuclear star cluster.
If the progenitor was a globular cluster, 
it would be the 
most metal-poor globular cluster known.
\end{abstract}



\section{Introduction}
\label{intro}

The stellar halo of the Milky Way is filled with structure.
Overdensities in physical, phase, and chemical space
reveal a rich history of accretion.
Reconstructing this history requires knowledge of the
characteristics of the stellar systems
and the Milky Way's mass distribution, so
it is important to identify the nature and properties of
the progenitor of new structure discovered
in the Milky Way halo.

\citet{ibata19a},
hereafter referred to as \citetalias{ibata19a},
discovered a number of stellar streams 
at high Galactic latitude
($|b| >$~20\degree)
and distance $d >$~1~kpc from the Sun.
Their search relied on the second data release from the 
\textit{Gaia} satellite to identify significant 
features in position, \textit{Gaia} photometry, and proper motions.
Cross-matching
with the Sloan Digital Sky Survey (SDSS; \citealt{yanny09}) and
LAMOST catalogs \citep{cui12}
revealed heliocentric radial velocity (\rv) and metallicity ([Fe/H])
estimates for a small subset of stars in the streams.

\citetalias{ibata19a}
identified 103 potential members of a stream
they named Sylgr.
It stretches across the
northern Galactic hemisphere between $-73^\circ \leq \ell \leq -95^\circ$
and $+50^\circ \leq b \leq +60$\degree.
They adopted a model for the Galactic potential and 
used an orbit integrator to derive a best-fit orbit for this stream.
The SDSS provided \rv\ measurements for 
three stars in the Sylgr stream, and these matched the predicted
\rv\ values at the longitude of the three stars.
This match boosts confidence in the existence of the stream and
confirms the membership of these three stars.
Sylgr has a highly radial, prograde rotating orbit,
with Galactic pericenter $2.49 \pm 0.02$~kpc
and apocenter $19.4 \pm 0.5$~kpc.
A color-magnitude diagram, constructed from \textit{Gaia}
broadband photometry, reveals that all of the candidate members
lie near or below the main sequence turnoff point, and
by construction these stars lie near a metal-poor 
12.5~Gyr isochrone.

At present, there is no obvious progenitor for the Sylgr stream,
but a few clues about the nature of the progenitor are available.
The STREAMFINDER algorithm, which \citetalias{ibata19a} used to
identify the Sylgr stream,
is best suited to detect thin and kinematically-cold streams
that may be formed by the disruption and dissolution of
globular clusters (GCs) \citep{malhan18a,malhan18c}.
The Sloan Extension for Galactic Understanding and Exploration (SEGUE) 
Stellar Parameter Pipeline (SSPP; \citealt{lee08}) 
estimated metallicities
for the three stars with SDSS spectroscopy,
[Fe/H] $= -$2.38, $-$2.58, and $-$3.10.
These values exhibit a considerable spread, and their average
is lower than that of all known GCs in the Milky Way 
\citep[$\feh > -2.4$;][]{carretta14ter8}.
Most of the other streams discovered by \citetalias{ibata19a}\
show minimal metallicity dispersion and higher mean metallicity
among their likely members,
providing the first hint that the progenitor of the Sylgr stream may not
have resembled a typical GC.~

Here we present high-resolution spectroscopy of two
high-probability members of the Sylgr stellar stream.
These spectra permit us to measure precise \rv\ values,
calculate stellar parameters, and 
derive detailed abundances for 13~elements for each star.
We present the new observational material,
including \rv\ measurements,
in Section~\ref{obs}.
We calculate stellar parameters and metallicities
in Section~\ref{params}.
We derive abundances of other elements
and compare them with various Galactic stellar populations
in Section~\ref{abundances}.
We discuss the implications of these results for the origin of the stream
in Section~\ref{discussion}
and summarize our conclusions
in Section~\ref{conclusions}.

\section{Observations}
\label{obs}

We observed  \jtwolong\ (hereafter \jtwo) and
\jeightlong\ (hereafter \jeight)
using the
Magellan Inamori Kyocera Echelle (MIKE; \citealt{bernstein03}) spectrograph
on the Landon Clay (Magellan~II) Telescope at 
Las Campanas Observatory, Chile.
MIKE is a double spectrograph, and
the 0\farcs7$\times$5\farcs0 
entrance slit and 2$\times$2 binning on the CCD yield a 
spectral resolving power of 
$R \equiv \lambda/\Delta\lambda \sim$~41,000 on the blue spectrograph
(3350 $< \lambda <$ 5000~\AA) and $R \sim$~36,000 on the red spectrograph
(5000 $< \lambda <$ 8300~\AA).
\jtwo\ and \jeight\ were each observed on 2019 April 11 and 12,
for total integration times of 5.00~hr and 5.61~hr, respectively.
Observations were made during excellent seeing conditions,
$\approx$~0\farcs4--0\farcs6.
The lunar contribution to the sky background was minimal for
\jtwo\ and non-existent for \jeight.

\begin{figure*}
\begin{center}
\includegraphics[angle=0,width=0.95\textwidth]{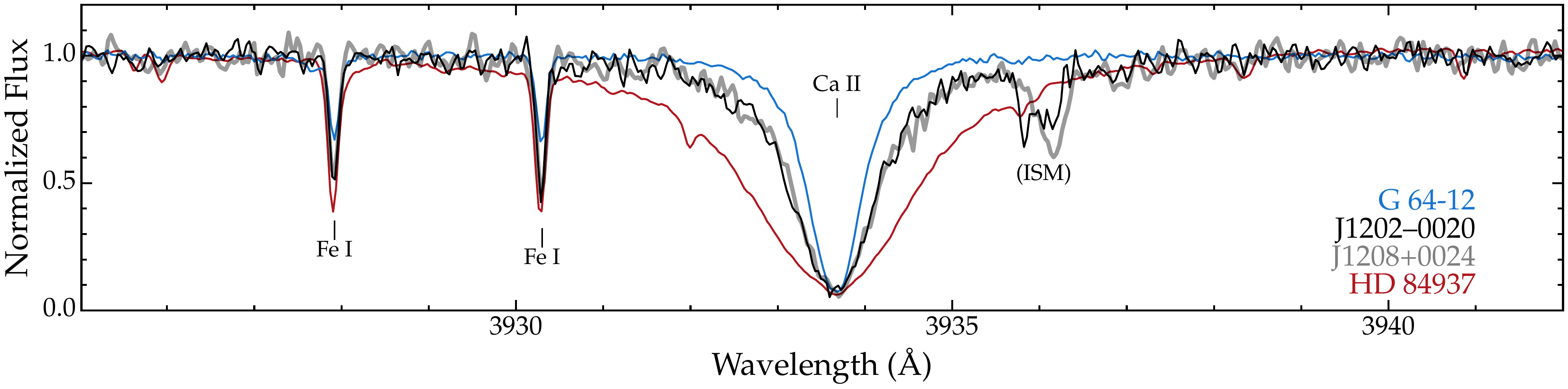} \\
\includegraphics[angle=0,width=0.95\textwidth]{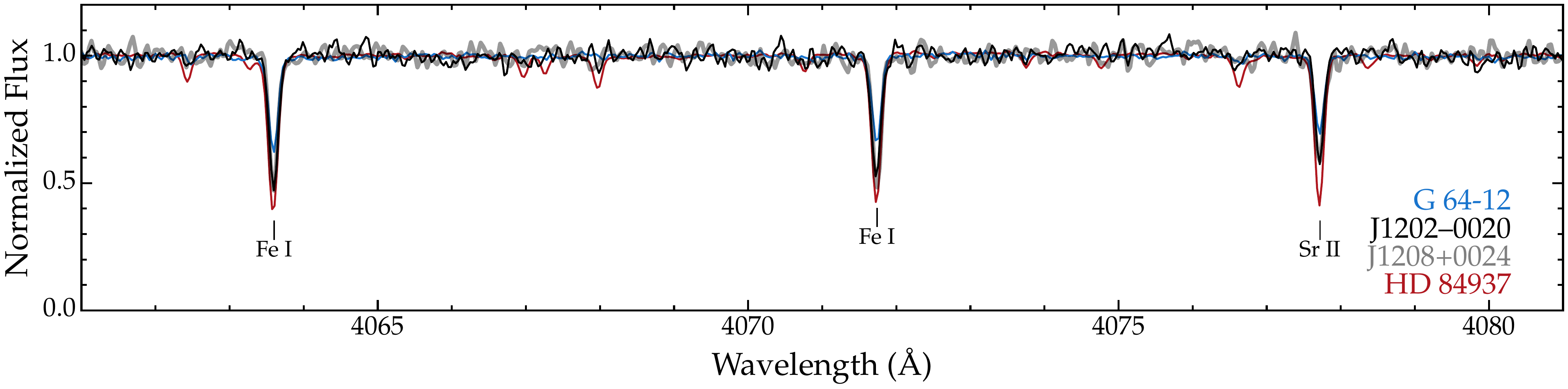} \\
\includegraphics[angle=0,width=0.95\textwidth]{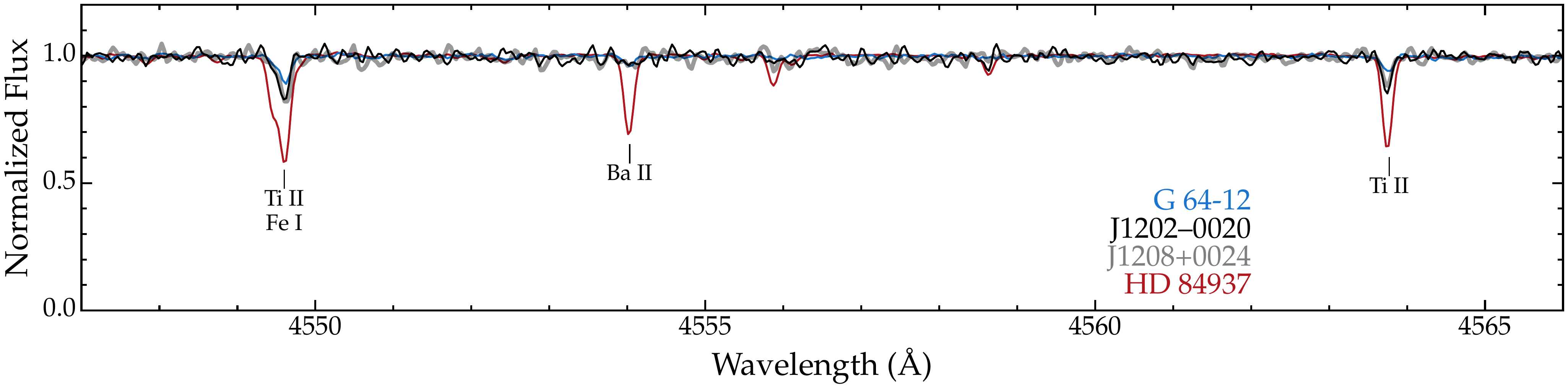} \\
\includegraphics[angle=0,width=0.95\textwidth]{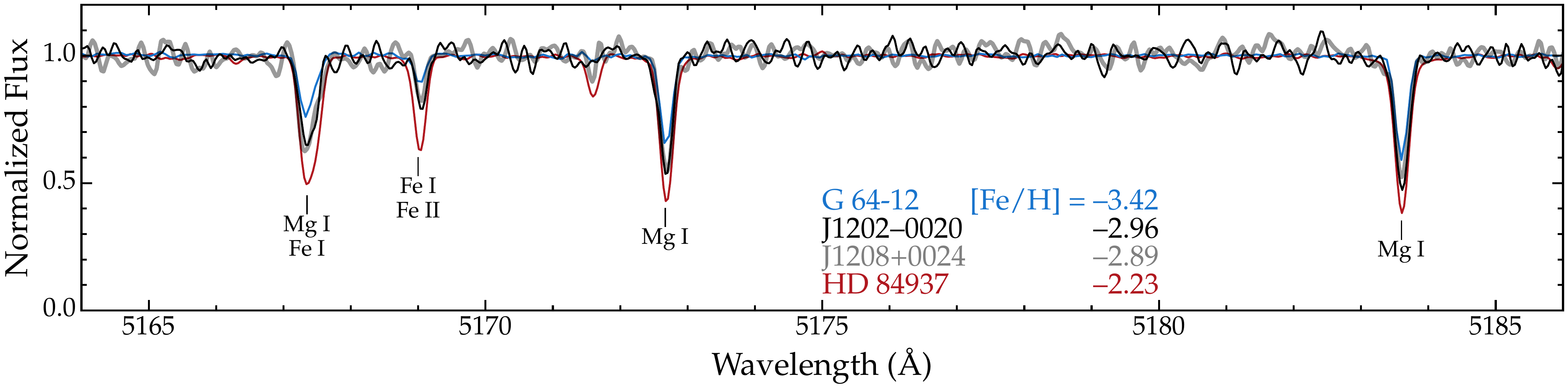} 
\end{center}
\vspace{-4mm}
\caption{
\label{specplot}
Selected regions of our MIKE spectra of \jtwo\ (thin black line)
and \jeight\ (bold gray line).
Lines of interest are marked.
Two other stars, also observed using the same MIKE setup,
are shown for comparison.
The spectrum of \gsix\ was presented in \citet{roederer14c},
and the spectrum of \hdeight\ is described in Section~\ref{obs}.
These two stars are only $\approx$~200~K warmer
than our target stars, and they
bracket the metallicities of our target stars,
as indicated in the bottom panel
and demonstrated by the relative strengths of 
metal lines.
 }
\end{figure*}

We use the CarPy MIKE reduction pipeline
\citep{kelson00,kelson03}
to perform the overscan subtraction, pixel-to-pixel flat field division,
image coaddition, cosmic ray removal, sky and scattered-light subtraction,
rectification of the tilted slit profiles along the orders,
spectrum extraction, and wavelength calibration.
We use IRAF \citep{tody93} to stitch together and
continuum-normalize the spectra.
Signal-to-noise ratios in the continuum range from
$\sim$~40/1~pix$^{-1}$ near 3950~\AA,
$\sim$~60/1~pix$^{-1}$ near 4550~\AA,
$\sim$~45/1~pix$^{-1}$ near 5200~\AA,
to 
$\sim$~70/1~pix$^{-1}$ near 6750~\AA\
for the final co-added spectrum of \jtwo,
and they are roughly 10\% lower for \jeight.
Figure~\ref{specplot} illustrates several regions
of the spectra around lines of interest.

We measure heliocentric \rv\ values by cross-correlating the
echelle order containing the Mg~\textsc{i} \textit{b} triplet 
against a metal-poor template spectrum obtained with MIKE.
We compute heliocentric velocity corrections using the 
IRAF ``rvcorrect'' task.
\citet{roederer14c} estimated uncertainties of $\approx$~0.7~\kmsec\
for \rv\ measured by this method.
The velocities we measure for \jtwo\ and \jeight, 
$-$203.0~\kmsec\ and $-$209.5~\kmsec, respectively,
compare well with the values measured from the SDSS spectra 
by the SSPP,
$-208.56 \pm 5.12$~\kmsec\ and
$-205.63 \pm 3.28$~\kmsec, respectively.
We also observed two velocity standard stars,
\hdeight\ and 
\object[HD 126587]{HD~126587}, on these nights using the same MIKE setup.
The heliocentric \rv\ values we measure,
$-$14.2~\kmsec\ and $+$150.0~\kmsec, respectively,
are in excellent agreement with previous measurements by
\citet{smith98}, \citet{ryan99}, \citet{carney01,carney03}, and
\citet{roederer14c}.
These comparisons suggest that the \rv\ zeropoint
of our measurements is reliable to 0.5~\kmsec\ or better.
We detect no significant \rv\ variations 
in our own measurements (separated by 1 day) or between our 
measurements and those from the SDSS spectra
(separated by 12~yr and 10~yr, respectively).
Our measurements support the model for the Sylgr stream
presented by \citetalias{ibata19a}.

\section{Stellar Parameters}
\label{params}

\subsection{Effective Temperature}
\label{temperature}

We calculate effective temperatures (\teff) using the 
metallicity-dependent color-\teff\ relations presented by
\citet{casagrande10}.
We first convert the SDSS $ugri$ photometry into Johnson-Cousins
$BVR_{c}I_{c}$ using the transformations presented by
\citet{jordi06}.
\citeauthor{casagrande10}\ present 
three color-\teff\ relations constructed from these bands:\
$B-V$, $V-R_{c}$, and $V-I_{c}$.
Near-infrared Two Micron All-Sky Survey (2MASS) photometry
is only available for one of the two stars,
and those $JHK$ magnitudes have uncertainties in excess
of 0.1~mag, so we do not make use of them.
We estimate the reddening, $E(B-V)$, by two methods:\ 
the dust maps presented by \citet{schlafly11}, 
and the amount of interstellar Na~\textsc{i} \textit{D} absorption 
\citep{bohlin78,spitzer78,ferlet85} as described in \citet{roederer18b}.
The reddening at these high Galactic latitudes is small in 
both cases, 
$E(B-V) \approx$~0.022 or less.
We adopt an initial model metallicity estimate of [M/H]~$= -3.0 \pm 0.5$
for both stars.
Table~\ref{phottab} presents the SDSS photometry,
calculated Johnson-Cousins photometry,
reddening estimates, 
and calculated \teff\ values from each of 
the three colors.
Table~\ref{phottab} also presents the \teff\ values we adopt,
which are weighted means of the predictions from
the three colors.
Using the \citet{ramirez05b} color-\teff\ relations,
rather than those from \citet{casagrande10},
would have yielded \teff\ values higher by 74~K on average.
The \citet{alonso99b} color-\teff\ relations
would have yielded \teff\ values cooler by 
250~K on average.

\begin{deluxetable*}{lccccccccc}
\tablecaption{Radial Velocities, Photometry, Calculated \teff\ Values, and Adopted Model Atmosphere Parameters
\label{phottab}}
\tablewidth{0pt}
\tabletypesize{\scriptsize}
\tablehead{
\colhead{Star} &
\colhead{\rv} &
\colhead{$u$} &
\colhead{$B$} &
\colhead{$E(B-V)$} &
\colhead{\teff($B-V$)} &
\colhead{Adopted} &
\colhead{Adopted} &
\colhead{Adopted} &
\colhead{Adopted} \\
\colhead{} &
\colhead{} &
\colhead{$g$} &
\colhead{$V$} &
\colhead{} &
\colhead{\teff($V-R_{c}$)} &
\colhead{\teff} &
\colhead{\logg} &
\colhead{\vt} &
\colhead{[M/H]} \\
\colhead{} &
\colhead{} &
\colhead{$r$} &
\colhead{$R_{c}$} &
\colhead{} &
\colhead{\teff($V-I_{c}$)} &
\colhead{} &
\colhead{} &
\colhead{} &
\colhead{} \\
\colhead{} &
\colhead{} &
\colhead{$i$} & 
\colhead{$I_{c}$} &
\colhead{} &
\colhead{} &
\colhead{} &
\colhead{} &
\colhead{} &
\colhead{} \\
\colhead{} &
\colhead{(\kmsec)} &
\colhead{(mag)} & 
\colhead{(mag)} &
\colhead{(mag)} &
\colhead{(K)} &
\colhead{(K)} &
\colhead{[cgs]} &
\colhead{(\kmsec)} &
\colhead{(dex)} 
}
\startdata
\jtwolong    & $-$203.3 $\pm$ 0.7 & 18.44 $\pm$ 0.02 & 17.89 $\pm$ 0.02 & 0.020 $\pm$ 0.01 & 6206 $\pm$ 140 & 6200 $\pm$ 83 & 4.47 $\pm$ 0.2 & 1.15 $\pm$ 0.1 & $-$3.0 $\pm$ 0.1 \\
             &                    & 17.56 $\pm$ 0.01 & 17.45 $\pm$ 0.01 &                  & 6556 $\pm$ 198 &&&& \\
             &                    & 17.33 $\pm$ 0.01 & 17.19 $\pm$ 0.01 &                  & 6092 $\pm$ 107 &&&& \\
             &                    & 17.23 $\pm$ 0.01 & 16.82 $\pm$ 0.01 &                  &  \\
\hline
\jeightlong  & $-$209.5 $\pm$ 0.7 & 18.63 $\pm$ 0.01 & 18.09 $\pm$ 0.02 & 0.022 $\pm$ 0.01 & 6255 $\pm$ 142 & 6242 $\pm$ 84 & 4.46 $\pm$ 0.2 & 1.15 $\pm$ 0.1 & $-$3.0 $\pm$ 0.1 \\
             &                    & 17.77 $\pm$ 0.01 & 17.67 $\pm$ 0.01 &                  & 6613 $\pm$ 202 &&&& \\
             &                    & 17.55 $\pm$ 0.01 & 17.41 $\pm$ 0.01 &                  & 6128 $\pm$ 108 &&&& \\
             &                    & 17.45 $\pm$ 0.01 & 17.04 $\pm$ 0.01 &                  &  \\
\enddata      
\tablecomments{%
The $BVR_{c}I_{c}$ magnitudes are calculated from the 
SDSS $gri$ magnitudes using 
the Population~II star transformations of \citet{jordi06}.
}
\end{deluxetable*}

\subsection{Surface Gravity}
\label{gravity}

We calculate the log of the surface gravity,
\logg, by interpolating 12~Gyr,
$\alpha$-enhanced
Yale-Yonsei ($Y^{2}$) isochrones \citep{demarque04} 
in \teff.
We assume that both stars are dwarfs on the main sequence,
as suggested by the color-magnitude diagram for 
candidate stream members presented by \citetalias{ibata19a}.
We estimate uncertainties in \logg\ by varying \teff\ by its uncertainty,
[M/H] by $\pm 1.0$~dex,
and letting the age range from 8 to 13~Gyr.
These values are presented in Table~\ref{phottab}.
We also interpolate absolute magnitudes, $M_{V}$,
for \jtwo\ and \jeight\ from the $Y^{2}$ isochrones.
These values, 
$M_{V} = 5.04 \pm 0.28$ and $4.95 \pm 0.32$, respectively,
imply distances of 
$3.0 \pm 0.4$~kpc and $3.4 \pm 0.5$~kpc.

Parallax measurements provide an alternative method of
estimating the surface gravity.
Unfortunately, the best parallax measurements available at present,
from the second data release of the \textit{Gaia} mission
\citep{lindegren18}, 
are uncertain for the two stars in our sample
($\varpi = 0.06 \pm 0.13$~mas and 
$0.63 \pm 0.23$~mas for \jtwo\ and \jeight, respectively).
The \citetalias{ibata19a} model orbit predicts the
parallax for all stars, assuming a common distance as a function of
Galactic longitude (their figure~10).
The distance estimated at the longitude of these two stars
is approximately $3.7 \pm 0.4$~kpc, 
which is in agreement with distances
estimated from the $Y^{2}$ isochrones.

\subsection{Microturbulent Velocity and Model Metallicity}
\label{microturbulent}

We estimate the microturbulent velocity, \vt, and model metallicity, [M/H],
with the help of abundances derived from 
Fe~\textsc{i} and \textsc{ii} lines.
We measure equivalent widths (EWs) 
using a semi-automatic 
routine that fits Voigt or Gaussian line profiles to 
continuum-normalized spectra
\citep{roederer14c}.
We visually inspect each line, and
we discard any line that appears blended 
or otherwise compromised.
We examine a telluric spectrum simultaneously
with the stellar spectrum, and we also discard 
any lines that appear to be contaminated by
telluric absorption.
These Fe~\textsc{i} and \textsc{ii} lines are listed in 
Table~\ref{linetab}.

\begin{deluxetable*}{cccccccccc}
\tablecaption{Line List
\label{linetab}}
\tabletypesize{\small}
\tablehead{
\colhead{} &
\colhead{} &
\colhead{} &
\colhead{} &
\colhead{} &
\multicolumn{2}{c}{\jtwo} &
\colhead{} &
\multicolumn{2}{c}{\jeight} \\
\cline{6-7} 
\cline{9-10}
\colhead{Species} &
\colhead{$\lambda$} &
\colhead{E.P.} &
\colhead{\loggf} &
\colhead{Ref.} &
\colhead{EW} &
\colhead{$\log\epsilon$\tablenotemark{a}} &
\colhead{} &
\colhead{EW} &
\colhead{$\log\epsilon$\tablenotemark{a}} \\
\colhead{} &
\colhead{(\AA)} &
\colhead{(eV)} &
\colhead{} &
\colhead{} &
\colhead{(m\AA)} &
\colhead{} &
\colhead{} &
\colhead{(m\AA)} &
\colhead{} 
}
\startdata
Li~\textsc{i}  & 6707.80 & 0.00 &    0.17 &  1 & \nodata &    2.08 && \nodata &    2.11 \\
O~\textsc{i}   & 7771.94 & 9.15 &    0.37 &  1 & \nodata &$<$ 7.50 && \nodata &$<$ 7.30 \\
Na~\textsc{i}  & 5889.95 & 0.00 &    0.11 &  1 &    89.1 &    3.67 &&    85.8 &    3.66 \\
Na~\textsc{i}  & 5895.92 & 0.00 & $-$0.19 &  1 &    62.7 &    3.54 &&    43.5 &    3.26 \\
\enddata
\tablecomments{%
The complete version of Table~\ref{linetab} is available in the online edition
of the journal.
A short version is included here to demonstrate its form and content.}
\tablenotetext{a}{%
LTE abundances}
\tablerefs{%
1 = \citet{kramida18};
2 = \citet{pehlivanrhodin17};
3 = \citet{aldenius09};
4 = \citet{lawler89}, using HFS from \citet{kurucz95};
5 = \citet{wood13};
6 = \citet{wood14v} for \loggf\ values and HFS;
7 = \citet{sobeck07};
8 = \citet{denhartog11} for both \loggf\ values and HFS;
9 = \citet{ruffoni14};
10 = \citet{belmonte17};
11 = \citet{lawler15} for \loggf\ values and HFS; 
12 = \citet{wood14ni};
13 = \citet{roederer12b};
14 = \citet{kramida18}, using HFS/IS from \citet{mcwilliam98};
15 = \citet{lawler01eu}, using HFS/IS from \citet{ivans06}.
}
\end{deluxetable*}

We make initial estimates for \vt\ (1.0~\kmsec) and [M/H] ($-$3.0)
and interpolate one-dimensional, 
hydrostatic model atmospheres from the $\alpha$-enhanced
ATLAS9 grid of models \citep{castelli04}
using an interpolation code provided by
A.\ McWilliam (2009, private communication).  
We derive Fe abundances 
using a recent version of the 
line analysis software MOOG
(\citealt{sneden73}; 2017 version), which
assumes local thermodynamic equilibrium (LTE).~
This version of MOOG treats
Rayleigh scattering, which affects 
the continuous opacity at shorter wavelengths,
as isotropic, coherent scattering
\citep{sobeck11}, 
although this update
has no effect ($<$~0.01~dex) on the abundances derived
for warm stars such as these
\citep{roederer18b}.
We adopt damping constants for collisional broadening
with neutral hydrogen from \citet{barklem00h}
and \citet{barklem05feii}, when available,
otherwise
we adopt the standard \citet{unsold55} recipe.
We adopt the \vt\ value for each model that 
minimizes the correlation between
the abundance derived from Fe~\textsc{i} lines
and line strength.
We adopt the [M/H] value for each model that
approximately matches the iron 
(Fe, $Z =$~26) abundance derived from Fe~\textsc{i} lines.
We iteratively determine \vt\ and [M/H], and the final 
model atmosphere parameters are reported in Table~\ref{phottab}.

We can only measure 2 or 3 weak
Fe~\textsc{ii} lines in our spectra of \jtwo\ and \jeight,
so we rely on Fe~\textsc{i} lines to set [M/H].~
Non-LTE corrections for a fair fraction of the Fe~\textsc{i} lines
measured in these two stars (16/43 and 15/48 lines, respectively) 
can be interpolated from the pre-computed grids presented 
in the INSPECT database by
\citet{bergemann12} and \citet{lind12}.
These corrections are small and consistent, with an average
non-LTE correction to the LTE abundances of 
$+$0.04~dex ($\sigma =$~0.01~dex).
The final non-LTE metallicities for \jtwo\ and \jeight\ are
[Fe/H]~$= -2.96 \pm 0.09$ and $-2.89 \pm 0.09$, 
respectively, and
we adopt [M/H]~$= -$3.0 for the model metallicity of each star.

\subsection{Comparison with Previous Results}
\label{previous}

The SSPP fit for \teff, \logg, and [Fe/H] for each of these stars:\
\teff\ $= 6452 \pm 61$~K,
\logg\ $= 3.77 \pm 0.18$, and 
[Fe/H] $= -2.38 \pm 0.11$ dex 
for \jtwo;
\teff\ $= 6502 \pm 30$~K,
\logg\ $= 3.60 \pm 0.15$, and 
[Fe/H] $= -3.10 \pm 0.08$ dex 
for \jeight.
These \teff\ values are warmer by about 250~K than the values we
derive from color-\teff\ relations.
We interpolate model atmospheres with these parameters 
and rederive the Fe abundance from Fe~\textsc{i} lines.
The SSPP model parameters
favor higher microturbulent velocities,
\vt\ $=$ 1.5~\kmsec, and 
they introduce a steep correlation
between the lower excitation potential (E.P.)\
and derived abundance,
which implies that the \teff\ values are too warm.
The warmer SSPP \teff\ value can account for
about half of the [Fe/H] discrepancy for \jtwo;
the remaining 0.3~dex discrepancy remains unexplained.
The \logg\ values from the SSPP
suggest that both stars are subgiants that have evolved beyond the 
main sequence turnoff.
The distances implied by this more luminous evolutionary state,
$d\approx 8 \pm 1$~kpc, are in  
conflict with the distances predicted by the \citetalias{ibata19a}
model orbit, $d=3.7 \pm 0.4$~kpc.
We conclude that both stars are on the main sequence,
and we adopt the stellar parameters derived in 
Sections~\ref{temperature}--\ref{microturbulent}.

\section{Abundance Analysis}
\label{abundances}

\subsection{Calculations}
\label{abundcalc}

We adopt the standard nomenclature 
for elemental abundances and ratios.
The abundance of element X is defined
as the number of X atoms per 10$^{12}$ H atoms,
$\log\epsilon$(X)~$\equiv \log_{10}(N_{\rm X}/N_{\rm H})+$12.0.
The ratio of the abundances of elements X and Y relative to the
Solar ratio is defined as
[X/Y] $\equiv \log_{10} (N_{\rm X}/N_{\rm Y}) - \log_{10} (N_{\rm X}/N_{\rm Y})_{\odot}$.
We adopt the Solar photospheric abundances of \citet{asplund09}.
By convention,
abundances or ratios denoted with the ionization state
indicate
the total elemental abundance derived from transitions of
that particular ionization state 
after Saha ionization corrections have been applied.

We measure EWs for lines of other species
and derive abundances
using the same procedure described in Section~\ref{params}.
These lines are also listed in Table~\ref{linetab}.
We derive upper limits on the abundance when no
lines of a particular element are detected.
Table~\ref{linetab} lists the atomic data for all lines investigated,
references for the \loggf\ values,
references for any hyperfine splitting structure (HFS) and 
isotope shifts (IS) 
used in the line component patterns for spectrum synthesis,
EW measurements,
abundances derived from the lines,
and 
upper limits.

We derive abundances of lines broadened by HFS and/or IS
by matching synthetic spectra,
computed using MOOG, to the observed spectrum.
We generate line lists for the synthetic spectra using the LINEMAKE code 
(C.\ Sneden, 2015, private communication),
which is based on the line lists of \citet{kurucz11}
and includes many updates based on 
experimental laboratory data.
We consider multiple isotopes in the syntheses of 
Li, C, N, Ba, and Eu, adopting
$^{7}$Li/$^{6}$Li~$=$~1000,
$^{12}$C/$^{13}$C~$=$~90,
$^{14}$N/$^{15}$N~$=$~1000,
and the rapid neutron-capture process
(\rpro) isotopic fractions from \citet{sneden08}
for Ba and Eu.

Table~\ref{abundtab} lists the mean abundances,
abundance ratios, and the uncertainties in these values.
We estimate uncertainties in the $\log\epsilon$ abundances
and [X/Fe] ratios by simultaneously 
resampling the stellar parameters, 
EWs (or approximations to the EWs in the case of lines
analyzed by spectrum synthesis matching), and \loggf\ values,
and we recompute the abundances from each resample
\citep{roederer18c}.
We repeat this procedure $10^{3}$ times, and the
16th and 84th percentiles
of the resulting distributions are reported as the 
1$\sigma$ uncertainties
in Table~\ref{abundtab}.

We detect no molecular features in the spectrum of either star.
We estimate upper limits on the C and N abundances
by comparing the observed spectra with synthetic spectra of
the CH $G$ band near 4290--4330~\AA\
and the CN band near 3875--3883~\AA.~

Numerous studies have quantitatively
assessed the impact of non-LTE effects
on abundances in warm, metal-poor dwarfs like the two stars
analyzed here.
Line-by-line non-LTE corrections to the LTE abundances 
are available for some species:\
$-$0.06~dex for Li~\textsc{i} \citep{lind09};
$-$0.12 to $-$0.28~dex for Na~\textsc{i} \citep{lind11};
$+$0.04 to $+$0.08~dex for Mg~\textsc{i} \citep{osorio15,osorio16};
and
$+$0.00~dex for Sr~\textsc{ii} \citep{bergemann12sr}.
We also make approximations for a few other species.
\citet{andrievsky08} estimated that non-LTE corrections to the
abundances derived from Al~\textsc{i} resonance lines 
are $\approx +$0.6~dex,
and we adopt this correction.
Non-LTE calculations by \citet{bergemann10} indicated that
Cr~\textsc{i} lines underestimate the Cr abundance by 
$\approx$~0.3 to 0.4~dex,
and \citet{roederer14c} found that abundances derived from
Cr~\textsc{i} lines underestimate those derived from Cr~\textsc{ii} lines
by a similar amount.
We apply a $+$0.35~dex correction to the LTE abundances 
derived from Cr~\textsc{i} lines.
\citet{bergemann08} found that 
the Mn~\textsc{i} resonance triplet
near 4030~\AA\ underestimates abundances by several tenths of a dex.
Similarly, \citet{sneden16} found that these lines yielded abundances
lower by 
$\approx$~0.3~dex relative to abundances derived from
Mn~\textsc{ii} lines and higher-excitation Mn~\textsc{i} lines.
\citet{bergemann08} attributed this discrepancy to non-LTE effects,
and we apply a $+$0.3~dex correction to our LTE abundances.

\begin{deluxetable*}{cccccccccccccc}
\tablecaption{Derived Abundances
\label{abundtab}}
\tablewidth{0pt}
\tabletypesize{\small}
\tablehead{
\colhead{} &
\multicolumn{6}{c}{\jtwo} &
\colhead{} &
\multicolumn{6}{c}{\jeight} \\
\cline{2-7}
\cline{9-14}
\colhead{Species} &
\colhead{$\log\epsilon$} &
\colhead{$\sigma$} &
\colhead{[X/Fe]} &
\colhead{[X/Fe]} &
\colhead{$\sigma$} &
\colhead{$N_{\rm lines}$} &
\colhead{} &
\colhead{$\log\epsilon$} &
\colhead{$\sigma$} &
\colhead{[X/Fe]} &
\colhead{[X/Fe]} &
\colhead{$\sigma$} &
\colhead{$N_{\rm lines}$} \\
\colhead{} &
\colhead{LTE} &
\colhead{} &
\colhead{LTE} &
\colhead{Non-LTE} &
\colhead{} &
\colhead{} &
\colhead{} &
\colhead{LTE} &
\colhead{} &
\colhead{LTE} &
\colhead{Non-LTE} &
\colhead{} &
\colhead{} 
}
\startdata
Li~\textsc{i}  &    2.08 &   0.09 & \nodata                  & 2.02\tablenotemark{a}    &    0.09 &     1 & &    2.11 &   0.11 & \nodata                  & 2.08\tablenotemark{a}    &    0.11 &  1 \\
C (CH)         &$<$ 6.50 &\nodata &$<+$1.03                  & \nodata                  & \nodata &\nodata& &$<$ 6.50 &\nodata &$<+$0.96                  & \nodata                  & \nodata &\nodata\\
N (CN)         &$<$ 8.00 &\nodata &$<+$3.13                  & \nodata                  & \nodata &\nodata& &$<$ 8.10 &\nodata &$<+$3.16                  & \nodata                  & \nodata &\nodata\\
O~\textsc{i}   &$<$ 7.50 &\nodata &$<+$1.77                  & \nodata                  & \nodata &     1 & &$<$ 7.30 &\nodata &$<+$1.50                  & \nodata                  & \nodata &  1 \\
Na~\textsc{i}  &    3.61 &   0.08 & $+$0.33                  & $+$0.10                  &    0.04 &     2 & &    3.46 &   0.08 & $+$0.11                  & $-$0.08                  &    0.05 &  2 \\
Mg~\textsc{i}  &    4.90 &   0.09 & $+$0.26                  & $+$0.33                  &    0.06 &     5 & &    4.91 &   0.09 & $+$0.20                  & $+$0.27                  &    0.07 &  4 \\
Al~\textsc{i}  &    3.05 &   0.08 & $-$0.44                  & $+$0.16                  &    0.05 &     2 & &    2.92 &   0.08 & $-$0.64                  & $-$0.04                  &    0.06 &  2 \\
Si~\textsc{i}  &    4.83 &   0.10 & $+$0.28                  & \nodata                  &    0.07 &     1 & &    4.85 &   0.11 & $+$0.23                  & \nodata                  &    0.07 &  1 \\
Ca~\textsc{i}  &    3.80 &   0.10 & $+$0.42                  & \nodata                  &    0.10 &     7 & &    3.85 &   0.09 & $+$0.40                  & \nodata                  &    0.10 &  6 \\
Sc~\textsc{ii} &    0.59 &   0.11 & $+$0.40                  & \nodata                  &    0.15 &     1 & &    0.47 &   0.11 & $+$0.21                  & \nodata                  &    0.13 &  1 \\
Ti~\textsc{ii} &    2.54 &   0.08 & $+$0.55                  & \nodata                  &    0.13 &    10 & &    2.43 &   0.08 & $+$0.37                  & \nodata                  &    0.12 &  9 \\
V~\textsc{ii}  &$<$ 2.30 &\nodata &$<+$1.33                  & \nodata                  & \nodata &     3 & &$<$ 2.30 &\nodata &$<+$1.26                  & \nodata                  & \nodata &  3 \\
Cr~\textsc{i}  &    2.62 &   0.11 & $-$0.06                  & $+$0.29                  &    0.05 &     3 & &    2.54 &   0.10 & $-$0.21                  & $+$0.14                  &    0.05 &  3 \\
Mn~\textsc{i}  &    1.97 &   0.11 & $-$0.50                  & $-$0.20                  &    0.08 &     3 & &    1.99 &   0.12 & $-$0.55                  & $-$0.25                  &    0.10 &  3 \\
Fe~\textsc{i}  &    4.50 &   0.09 & $-$3.00\tablenotemark{b} & $-$2.96\tablenotemark{b} &    0.09 &    43 & &    4.57 &   0.09 & $-$2.93\tablenotemark{b} & $-$2.89\tablenotemark{b} &    0.09 & 48 \\
Fe~\textsc{ii} &    4.70 &   0.14 & $-$2.80\tablenotemark{b} & \nodata                  &    0.14 &     2 & &    4.68 &   0.14 & $-$2.82\tablenotemark{b} & \nodata                  &    0.14 &  3 \\
Co~\textsc{i}  &$<$ 2.90 &\nodata &$<+$0.87                  & \nodata                  & \nodata &     2 & &$<$ 2.80 &\nodata &$<+$0.70                  & \nodata                  & \nodata &  2 \\
Ni~\textsc{i}  &    3.33 &   0.11 & $+$0.07                  & \nodata                  &    0.05 &     3 & &    3.38 &   0.12 & $+$0.05                  & \nodata                  &    0.06 &  3 \\
Zn~\textsc{i}  &$<$ 2.90 &\nodata &$<+$1.30                  & \nodata                  & \nodata &     1 & &$<$ 2.90 &\nodata &$<+$1.23                  & \nodata                  & \nodata &  1 \\
Sr~\textsc{ii} &    0.16 &   0.10 & $+$0.25                  & $+$0.25                  &    0.15 &     2 & &    0.18 &   0.10 & $+$0.20                  & $+$0.20                  &    0.15 &  2 \\
Ba~\textsc{ii} &$<-$1.20 &\nodata &$<-$0.42                  & \nodata                  & \nodata &     2 & &$<-$1.05 &\nodata &$<-$0.34                  & \nodata                  & \nodata &  2 \\
Eu~\textsc{ii} &$<-$0.40 &\nodata &$<+$2.04                  & \nodata                  & \nodata &     3 & &$<-$0.30 &\nodata &$<+$2.07                  & \nodata                  & \nodata &  3 \\
\enddata      
\tablenotetext{a}{$\log\epsilon$ notation}
\tablenotetext{b}{[Fe/H]}
\end{deluxetable*}

\subsection{Robust Metal Abundances}
\label{iron}

Figure~\ref{specplot} illustrates two key points about the
metal abundances.
First, spectral lines of metals including Mg, Ca, Ti, and Fe
are virtually indistinguishable in \jtwo\ and \jeight.
Given that the stellar parameters of these two stars are 
statistically identical, it follows that the
metal abundances are also statistically identical.
Secondly, the absorption line depths in these two stars are
intermediate between the two comparison stars,
\gsix\ and \hdeight.
These stars are only slightly warmer 
($6492 \pm 103$~K and $6418 \pm 117$~K, respectively)
than \jtwo\ and \jeight\
($6200 \pm 83$~K and $6242 \pm 84$~K, respectively),
and their surface gravities 
($4.18 \pm 0.21$ and $4.16 \pm 0.14$)
are only slightly lower than \jtwo\ and \jeight\
($4.47 \pm 0.2$ and $4.46 \pm 0.2$),
so their spectra provide a fair comparison
\citep{roederer18b}.
It is evident from Figure~\ref{specplot} that 
the metallicity of \gsix\
([Fe/H]~$= -3.42 \pm 0.08$) is considerably 
lower than that of
\jtwo \ or \jeight.
Similarly, the metallicity of \hdeight\
([Fe/H]~$= -2.23 \pm 0.07$) is considerably
higher than that of \jtwo\ or \jeight.
This comparison demonstrates that
the two stars in the Sylgr stream have 
extremely low metallicities.

We recompute metallicities from Fe~\textsc{i} lines
under a variety of assumptions about the stellar parameters,
and in all cases the [Fe/H] ratios are
statistically identical.
If we use a model atmosphere computed using
the SSPP values 
(which we disfavor; see Section~\ref{previous}),
then \jtwo\ and \jeight\ have LTE metallicities of
[Fe/H]~$= -2.80 \pm 0.09$ and
$-2.74 \pm 0.09$, respectively.
If we use the stellar parameters inferred from a different
set of transformations from $griz$ to $BVR_{C}I_{C}$ 
\citep{jester05},
which result in slightly lower temperatures and higher gravities,
then \jtwo\ and \jeight\ have LTE metallicities of
[Fe/H]~$= -3.07 \pm 0.18$ and
$-3.03 \pm 0.18$, respectively.
If we use model atmospheres interpolated from the
MARCS grid \citep{gustafsson08},
instead of the ATLAS9 grid,
then \jtwo\ and \jeight\ have unchanged LTE metallicities of
[Fe/H] $= -3.00 \pm 0.09$ and
$-2.93 \pm 0.09$, respectively.

We conclude from these tests 
that \jtwo\ and \jeight\ have
identical metallicities, 
and that these metallicities are extremely low,
[Fe/H]~$\approx -$2.9 or so.

\subsection{Lithium}
\label{lithium}

Lithium (Li, $Z =$~3) abundances have been derived previously in many
unevolved metal-poor main sequence stars, 
which are thought to preserve the natal Li abundances.
The surface convection zones are relatively shallow,
so they prevent Li from being mixed to deeper layers
where it can be destroyed.
Early studies discovered that warm, metal-poor
main sequence stars showed a near-uniform ``plateau'' value,
$\log\epsilon$(Li)~$\approx$~2.05
\citep{spite82,spite84}.
More recent studies that included larger samples
of stars with [Fe/H]~$< -$2.5
found an average decrease of a few tenths of a dex in $\log\epsilon$(Li)
among the most metal-poor stars,
and some also found increased abundance scatter
(e.g., \citealt{ryan99,boesgaard05,aoki09li,melendez10}).

\begin{figure}
\includegraphics[angle=0,width=3.35in]{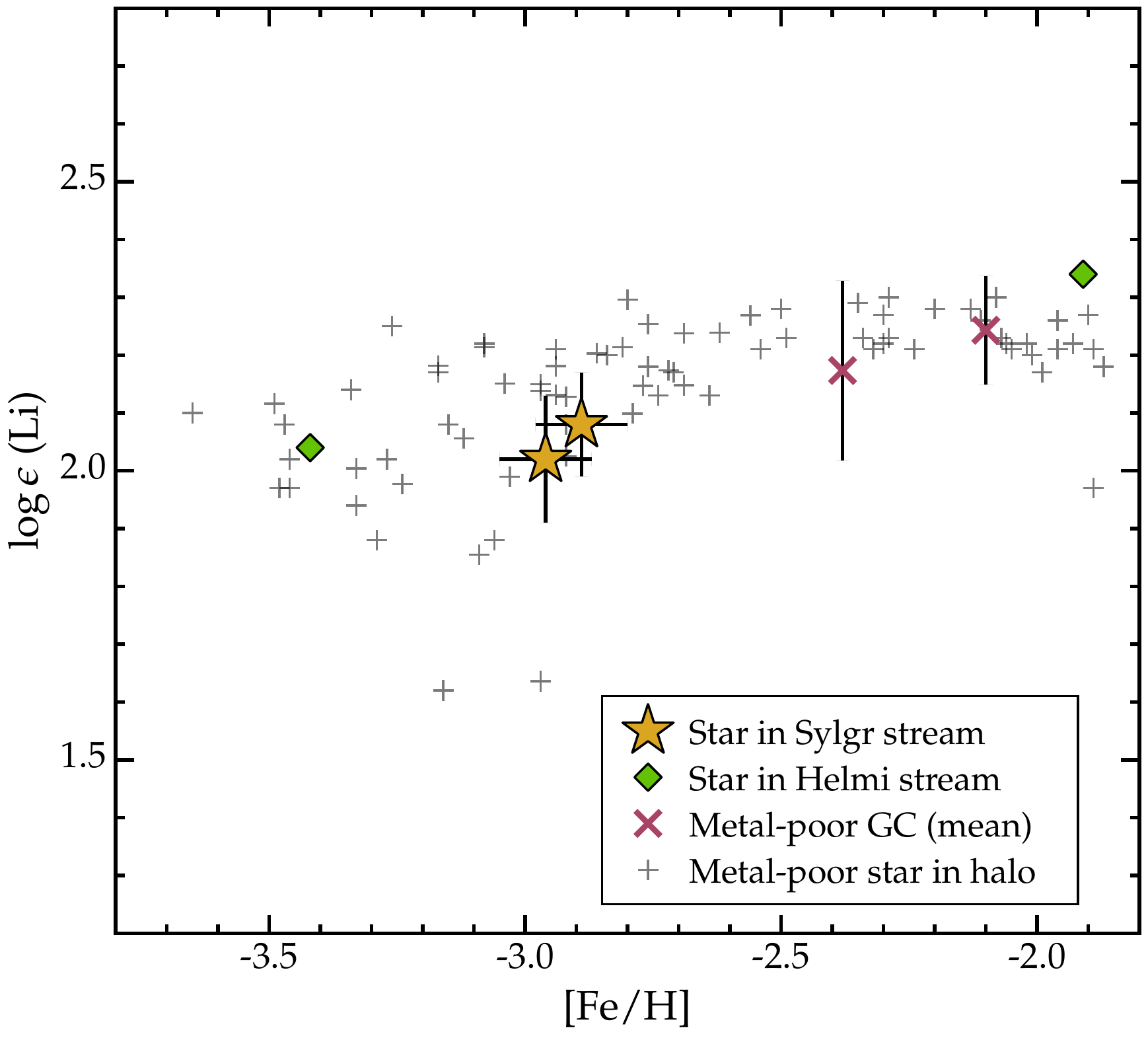}
\caption{
\label{lifeplot}
Comparison of $\log\epsilon$(Li) abundances in the
Sylgr stream stars with those in dwarf
(\teff~$>$~5700~K; \logg~$>$~3.8) 
stars in other metal-poor populations.
The halo field star samples are drawn from
\citet{boesgaard05},
\citet{asplund06},
\citet{bonifacio07,bonifacio12},
\citet{melendez10}, 
and
\citet{sbordone10},
all of whom applied non-LTE corrections to their Li abundances.
Duplications have been removed.
There are two GCs with [Fe/H]~$< -$2
for which Li abundances have been derived in dwarf stars.
The abundances for 
\object[M30]{M30} 
([Fe/H]~$= -$2.38, \citealt{cohen11n2419})
and
\object[NGC 6397]{NGC~6397}
([Fe/H]~$= -$2.10, \citealt{koch11})
are shown as the mean [Fe/H] ratios for each cluster
and the mean $\pm$ one standard deviation for Li
\citep{lind09n6397,gruyters16}.
The Li abundances for dwarf
stars in the stream discovered by \citet{helmi99}
are quoted from \citet{roederer14c}.
Li abundances have not been presented in the literature for
dwarf stars in any dSph or UFD galaxy.
 }
\end{figure}

Figure~\ref{lifeplot} compares the Li abundances of \jtwo\ and \jeight\ 
with three previous studies of Li abundances in 
warm, unevolved metal-poor dwarf stars
in the halo field (76~unique stars),
two GCs with [Fe/H]~$< -$2
(59~stars in \object[M30]{M30} and
174~stars in \object[NGC 6397]{NGC~6397}),
and two field stars associated with the 
stellar stream discovered by \citet{helmi99}.
It is clear from Figure~\ref{lifeplot}
that the Li abundances in these
two dwarf stars in the Sylgr stream
are broadly consistent with abundances
in other dwarf stars
with similar metallicity 
in other environments.

\subsection{Carbon, Nitrogen, and Oxygen}
\label{cno}

We do not detect any atomic or molecular lines of 
carbon (C, $Z =$~6), 
nitrogen (N, $Z =$~7), or
oxygen (O, $Z =$~8) in either star.
The upper limits derived from the non-detection of the
CH \textit{G} band near 4300~\AA\ indicate that the
[C/Fe] ratios are not enormously enhanced relative 
to the Solar ratio, but we cannot exclude 
[C/Fe]~$< +$1.0 or so in either star.
The non-detection of the CN bandhead near 3883~\AA\
yields only uninformative upper limits on the N abundances,
[N/Fe]~$< +$3.2 or so.
The non-detection of the O~\textsc{i} triplet near
7770~\AA\ also provides uninformative upper limits on the O abundances,
[O/Fe]~$< +$1.8 and $< +$1.5 in \jtwo\ and \jeight, respectively.

\subsection{Sodium through Nickel}
\label{alphairon}

\begin{figure*}
\begin{center}
\includegraphics[angle=0,width=2.27in]{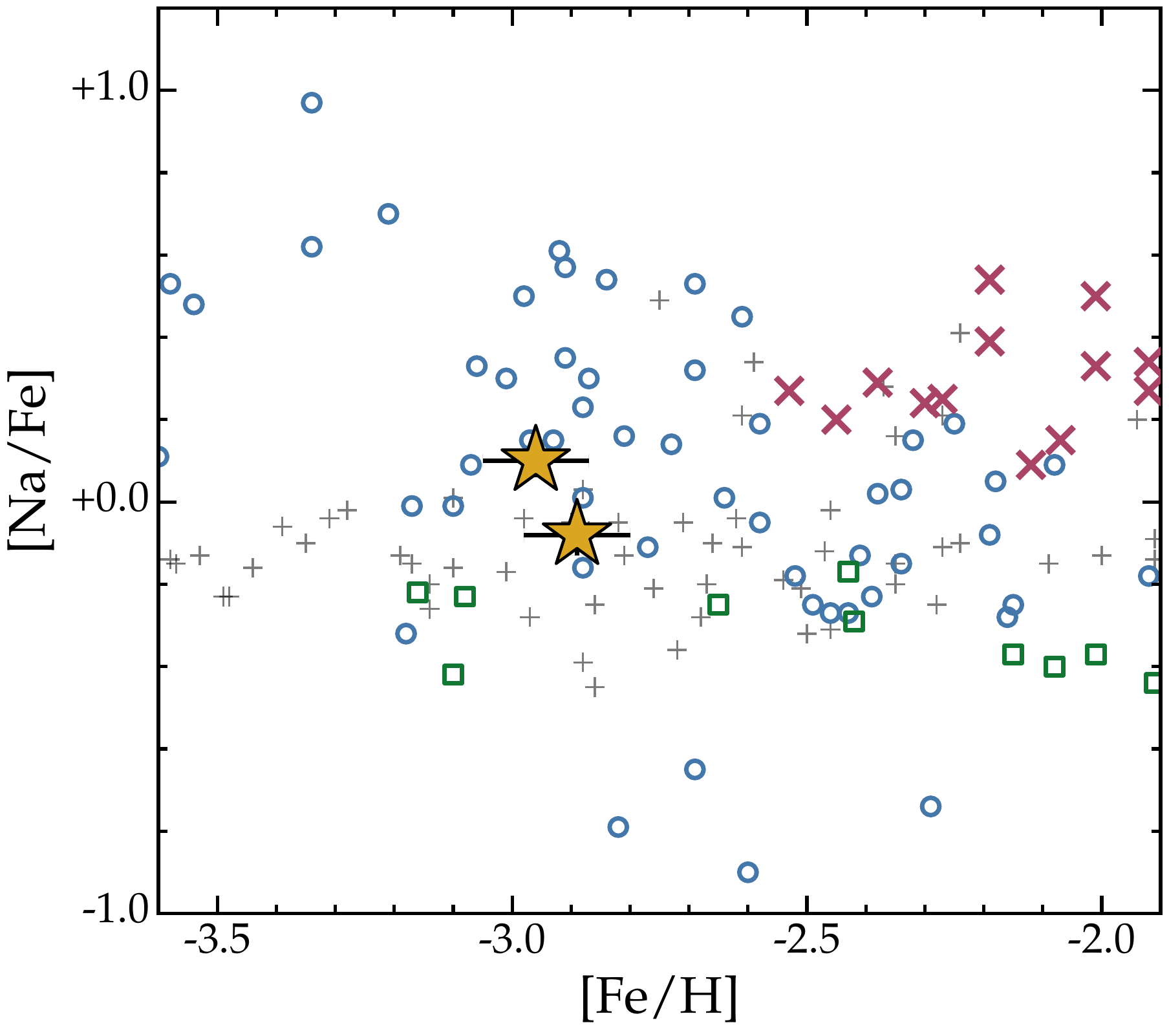}
\hspace*{0.02in}
\includegraphics[angle=0,width=2.27in]{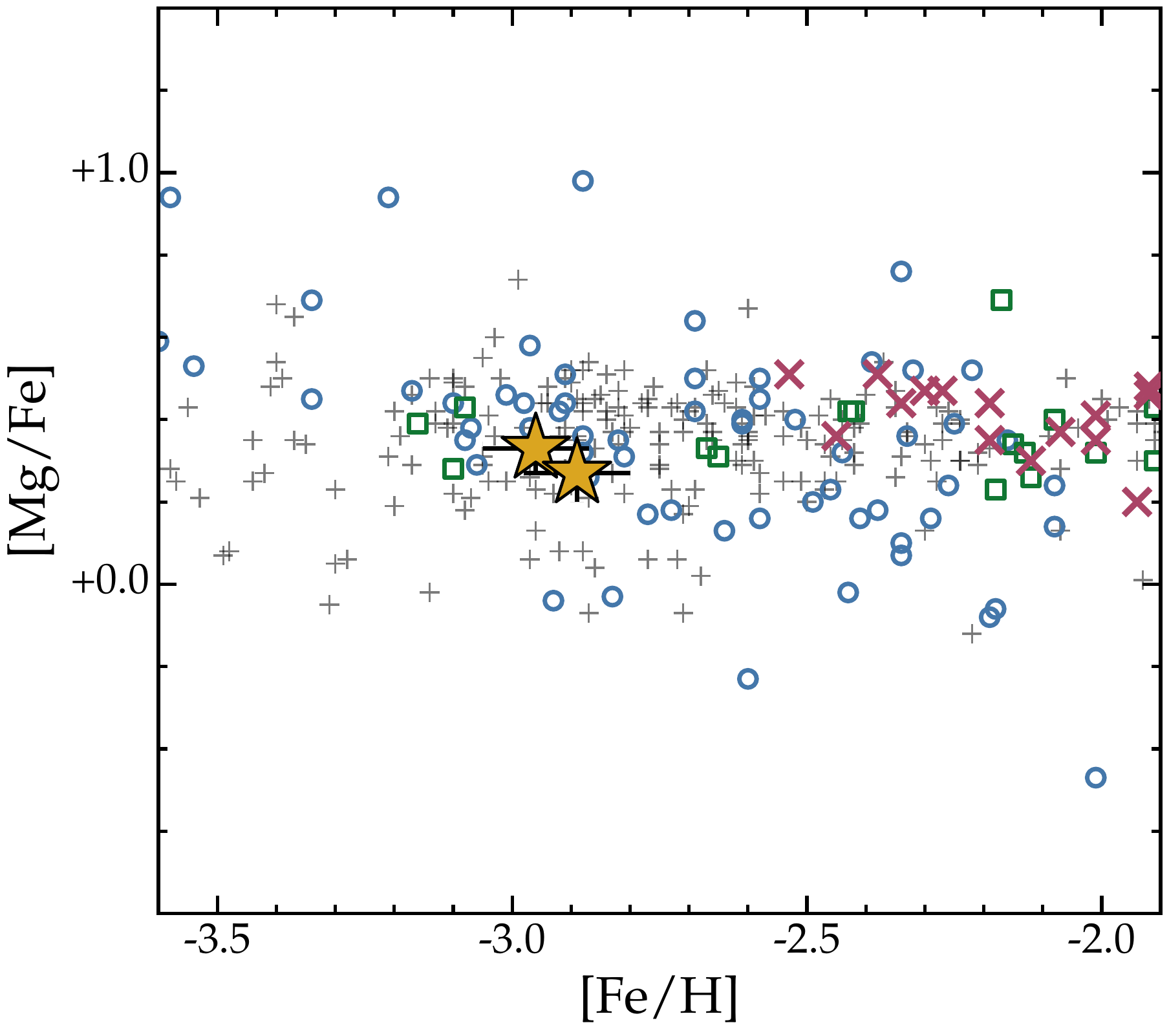} 
\hspace*{0.02in}
\includegraphics[angle=0,width=2.27in]{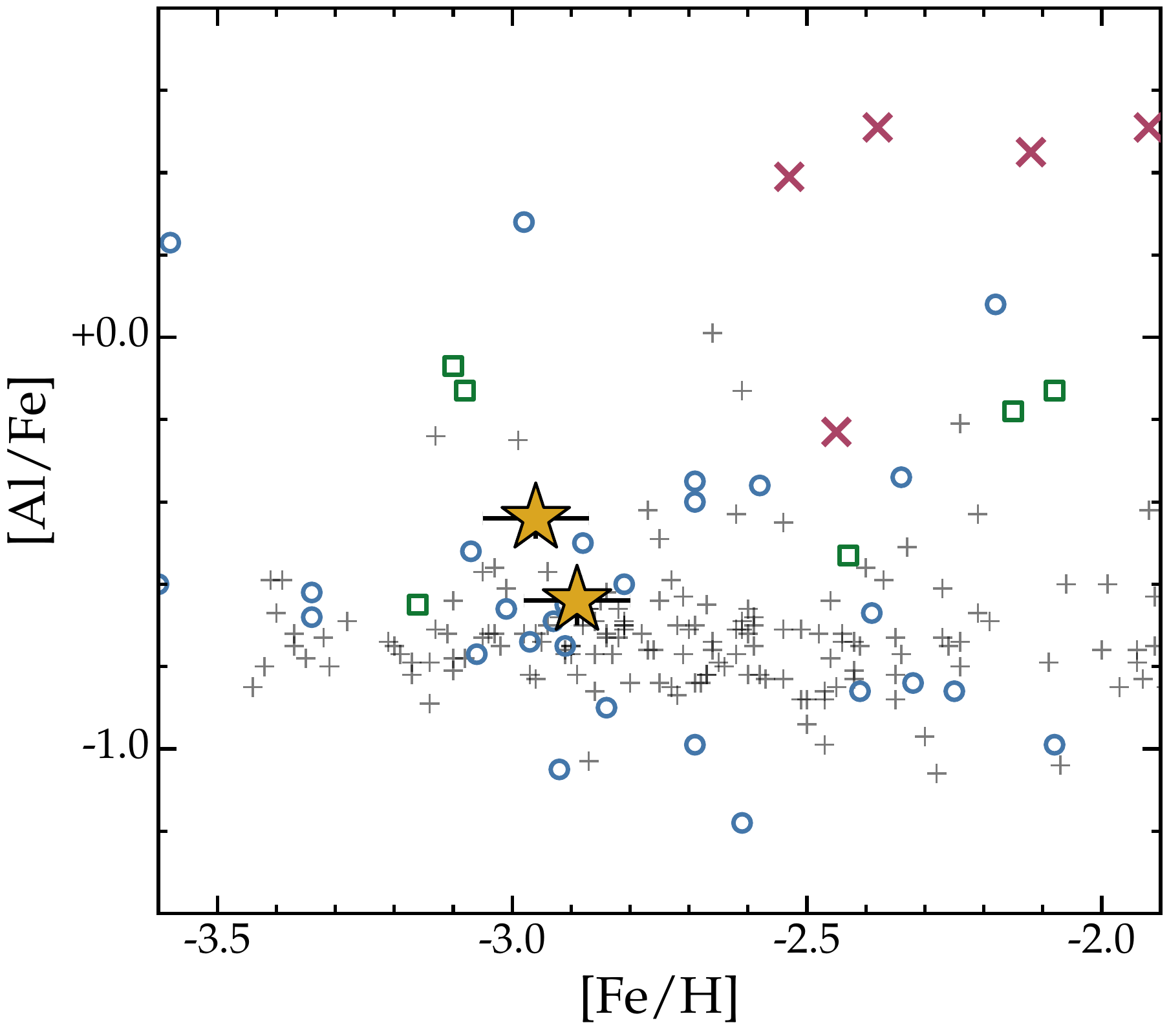} \\
\vspace*{-0.231in}
\includegraphics[angle=0,width=2.27in]{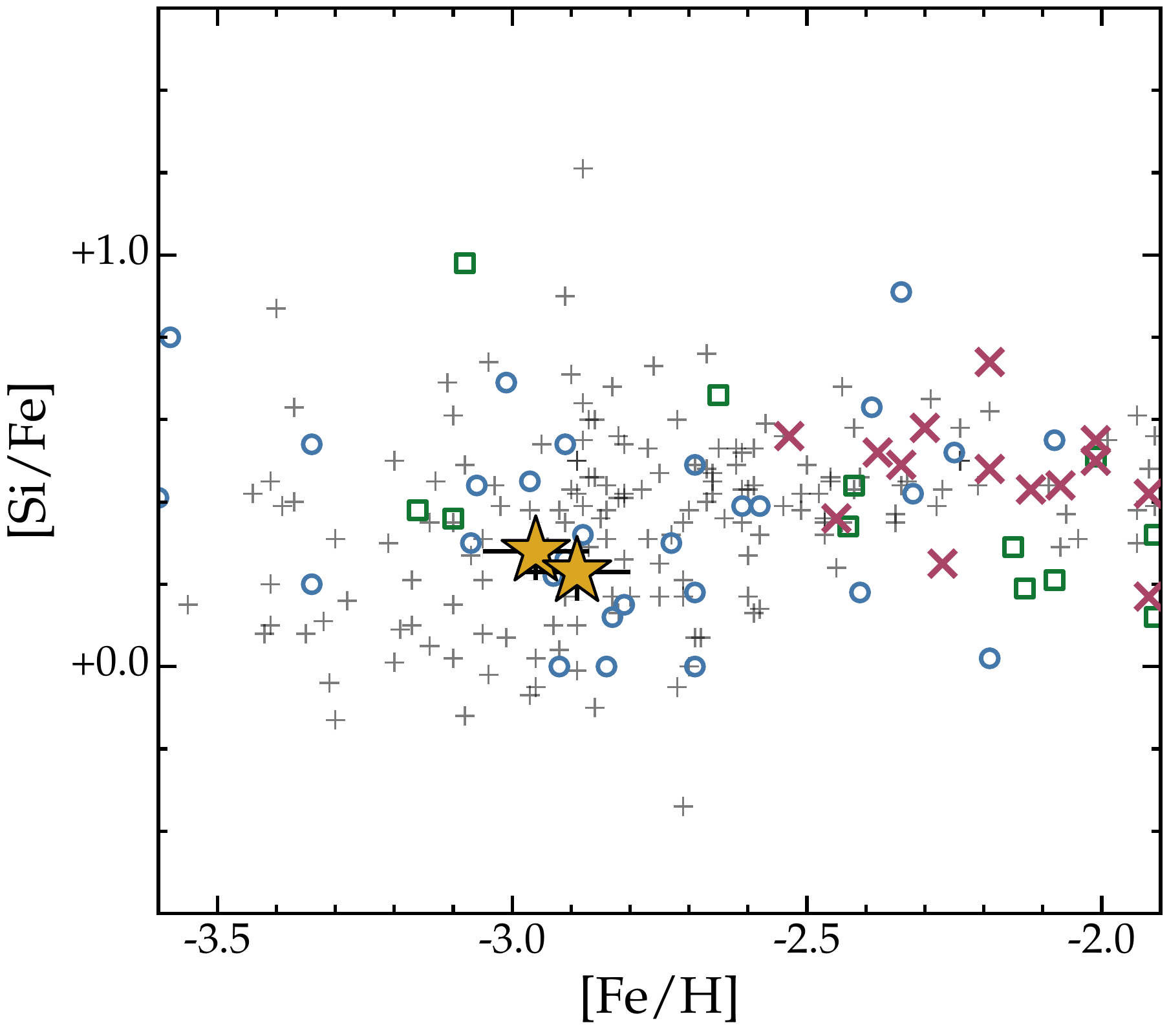}
\hspace*{0.02in}
\includegraphics[angle=0,width=2.27in]{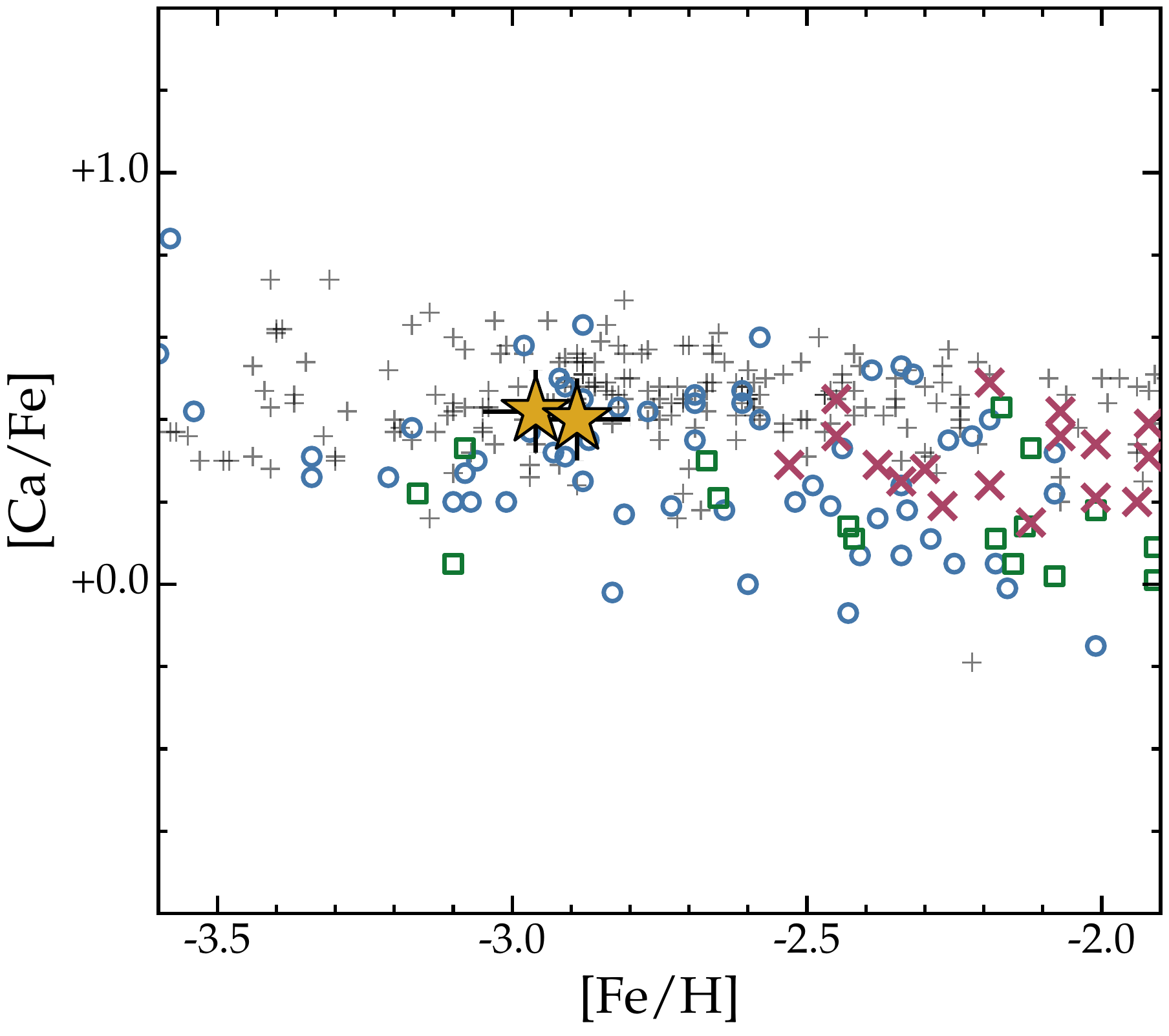} 
\hspace*{0.02in}
\includegraphics[angle=0,width=2.27in]{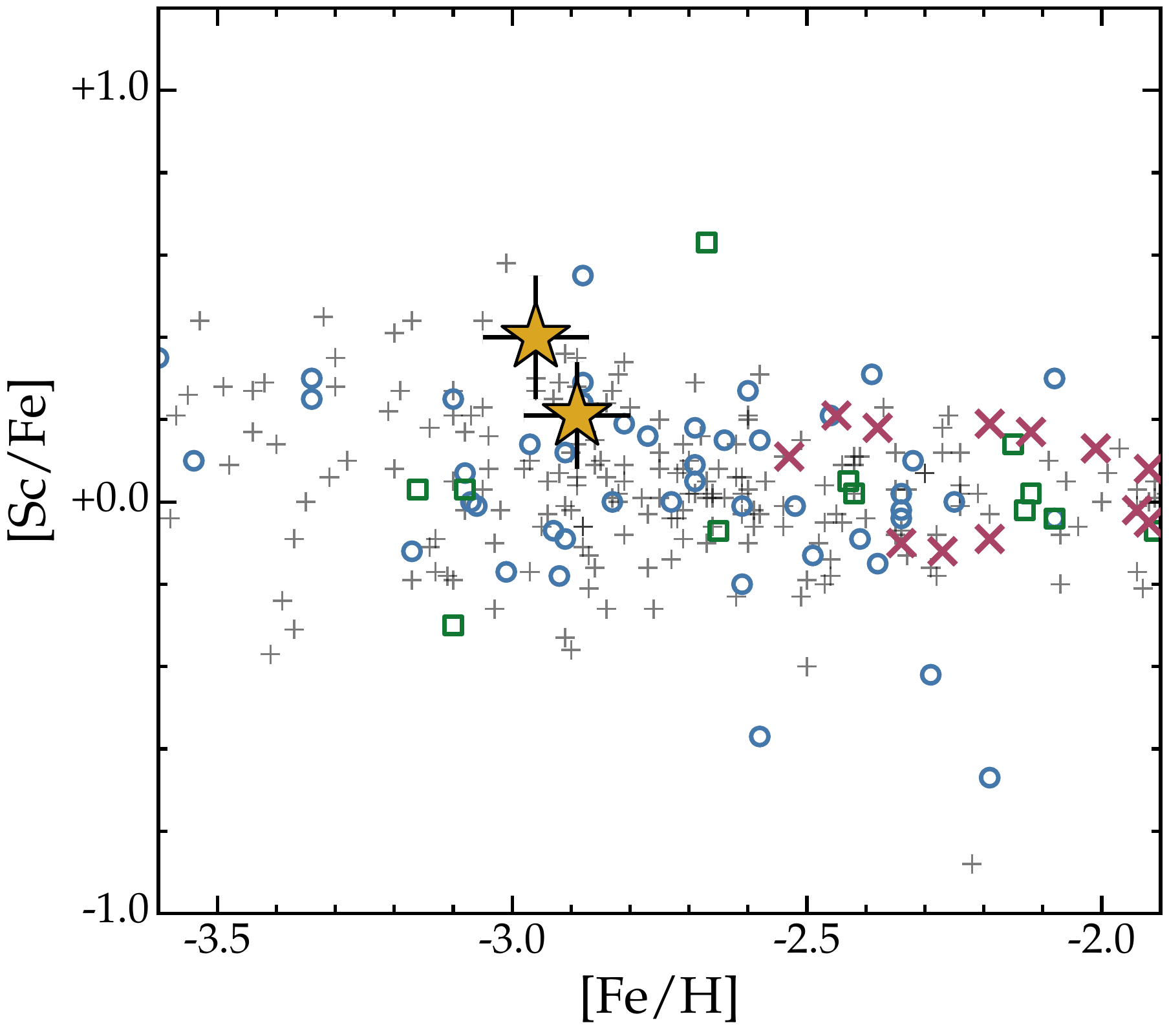} \\
\vspace*{-0.231in}
\includegraphics[angle=0,width=2.27in]{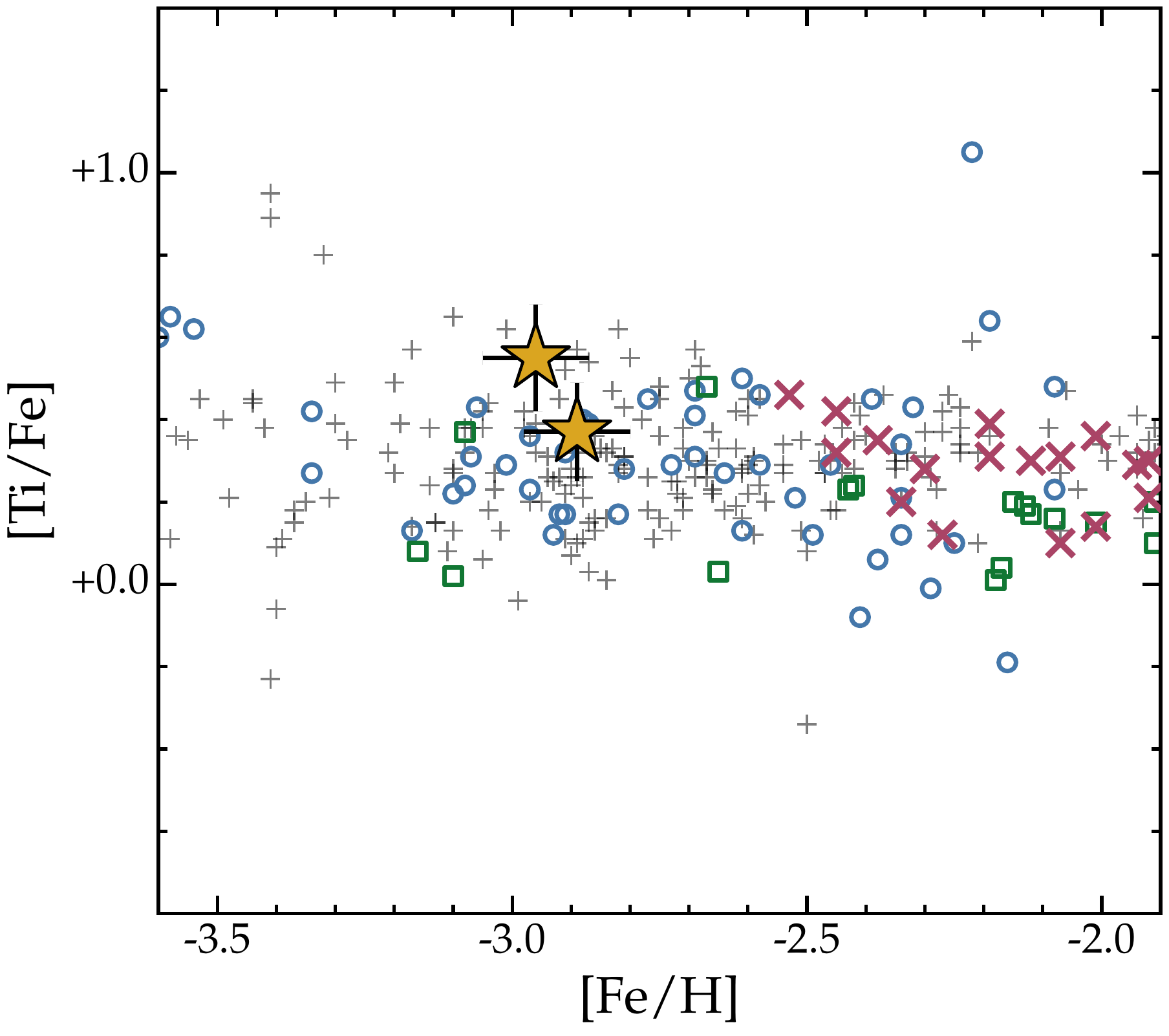}
\hspace*{0.02in}
\includegraphics[angle=0,width=2.27in]{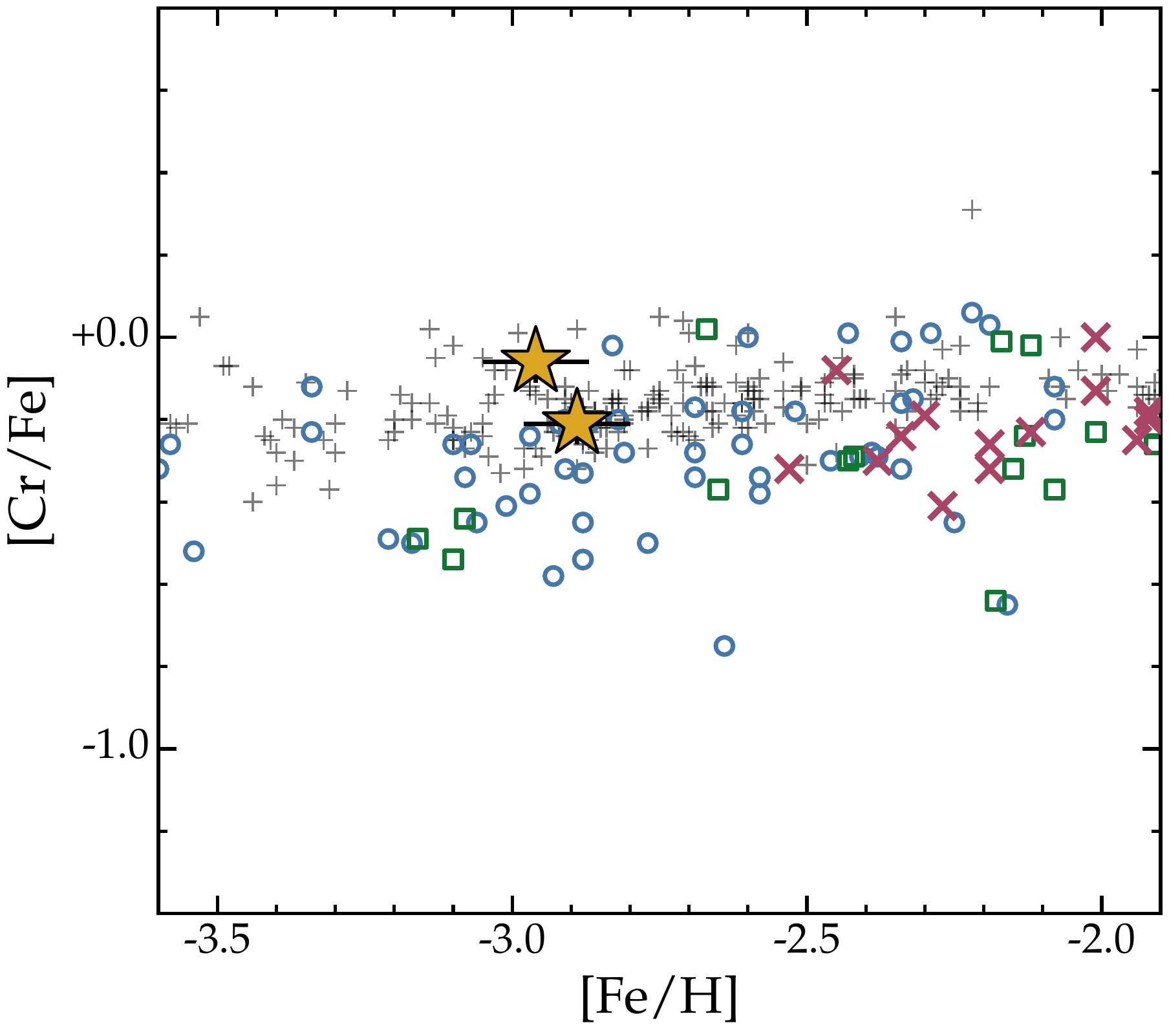} 
\hspace*{0.02in}
\includegraphics[angle=0,width=2.27in]{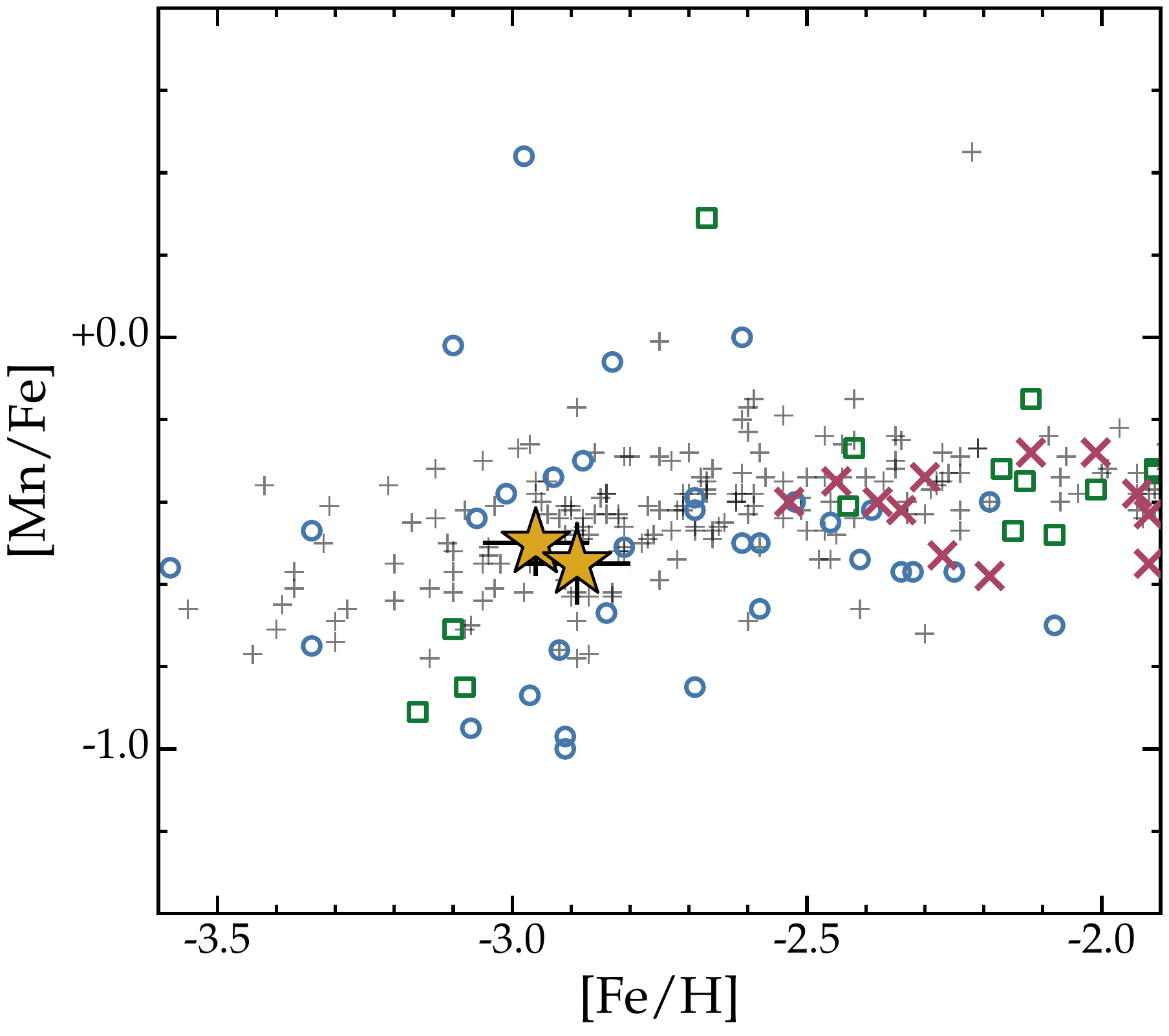} \\
\vspace*{-0.231in}
\includegraphics[angle=0,width=2.27in]{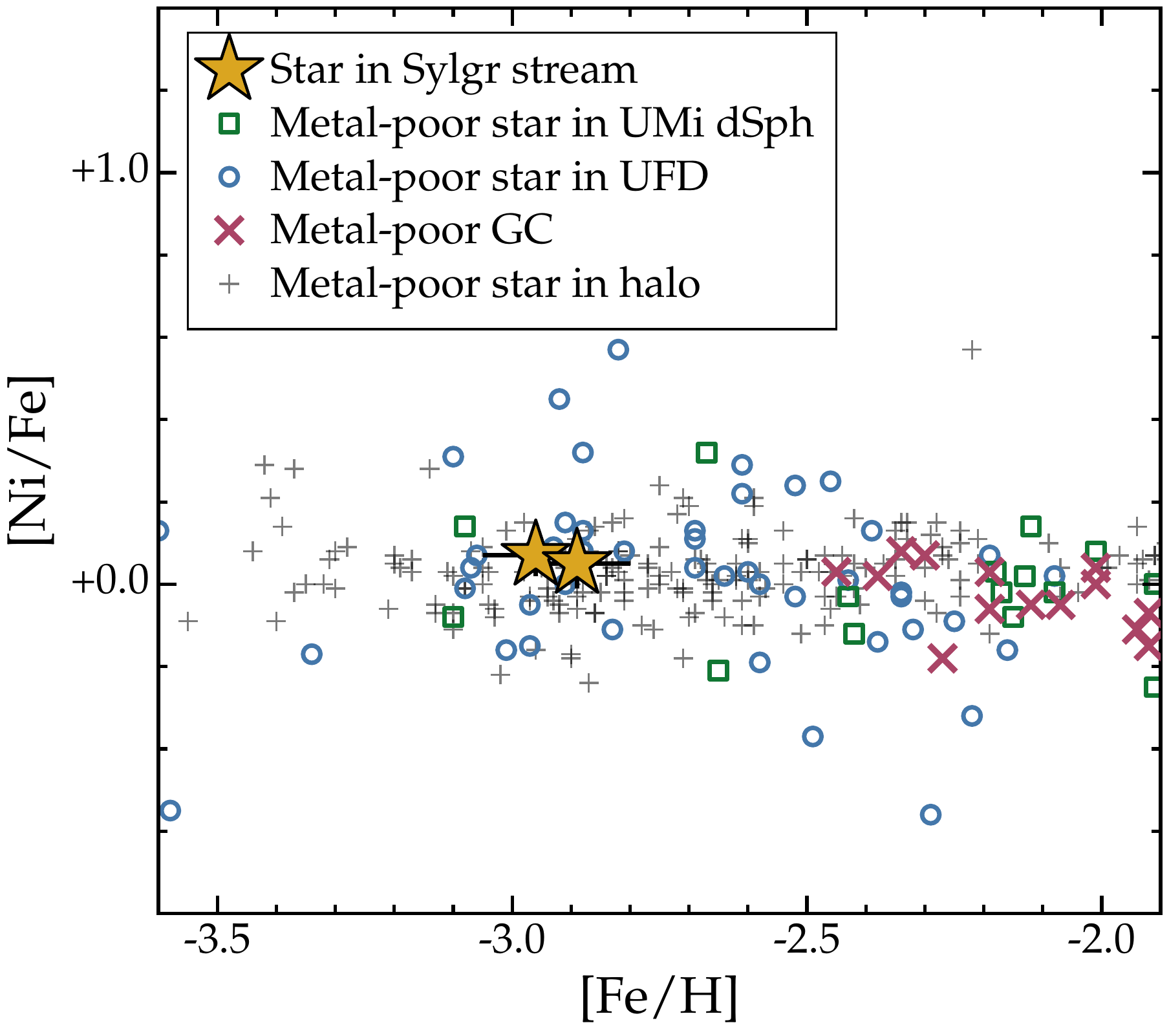}
\hspace*{0.02in}
\includegraphics[angle=0,width=2.27in]{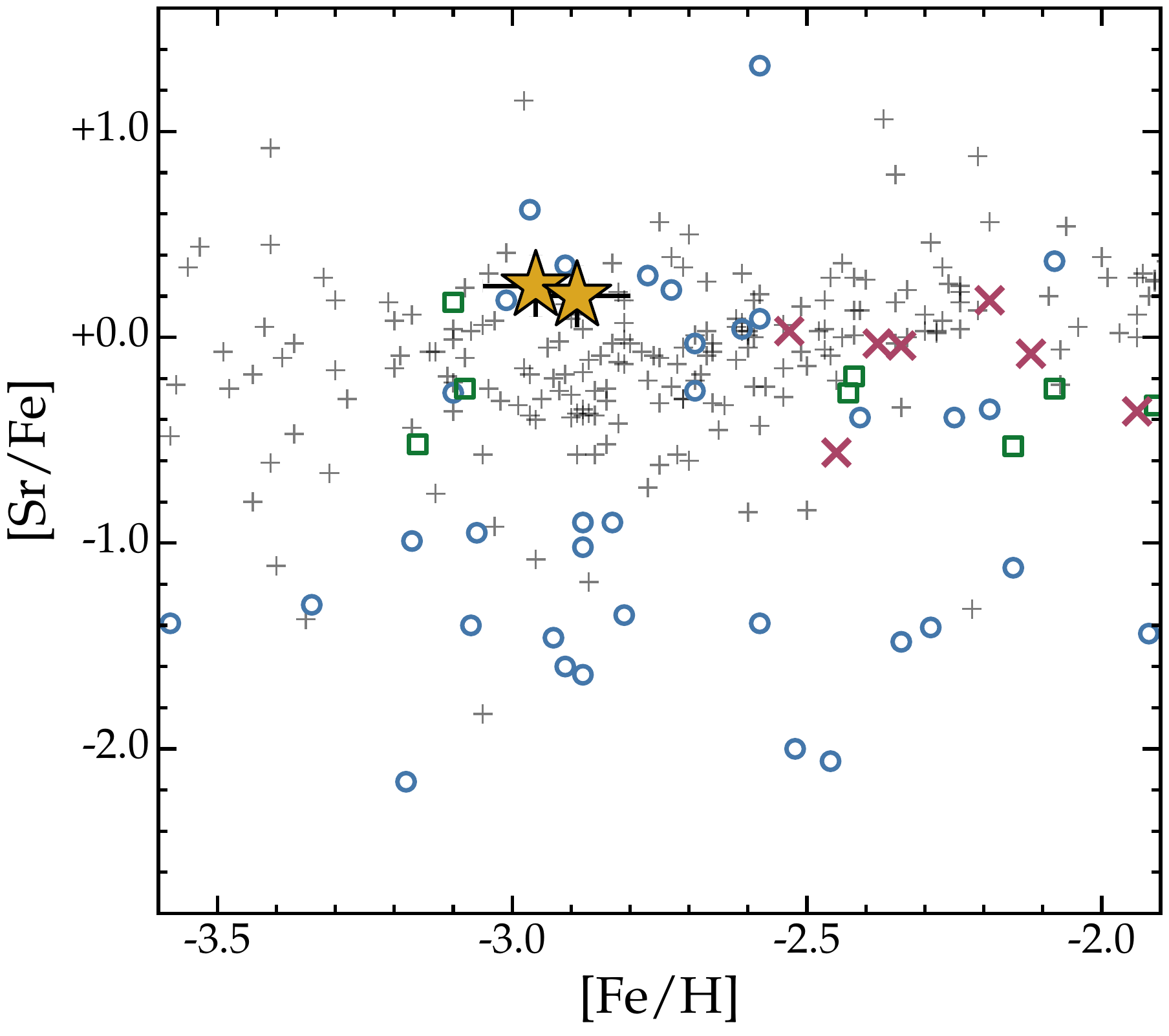} 
\hspace*{0.02in}
\includegraphics[angle=0,width=2.27in]{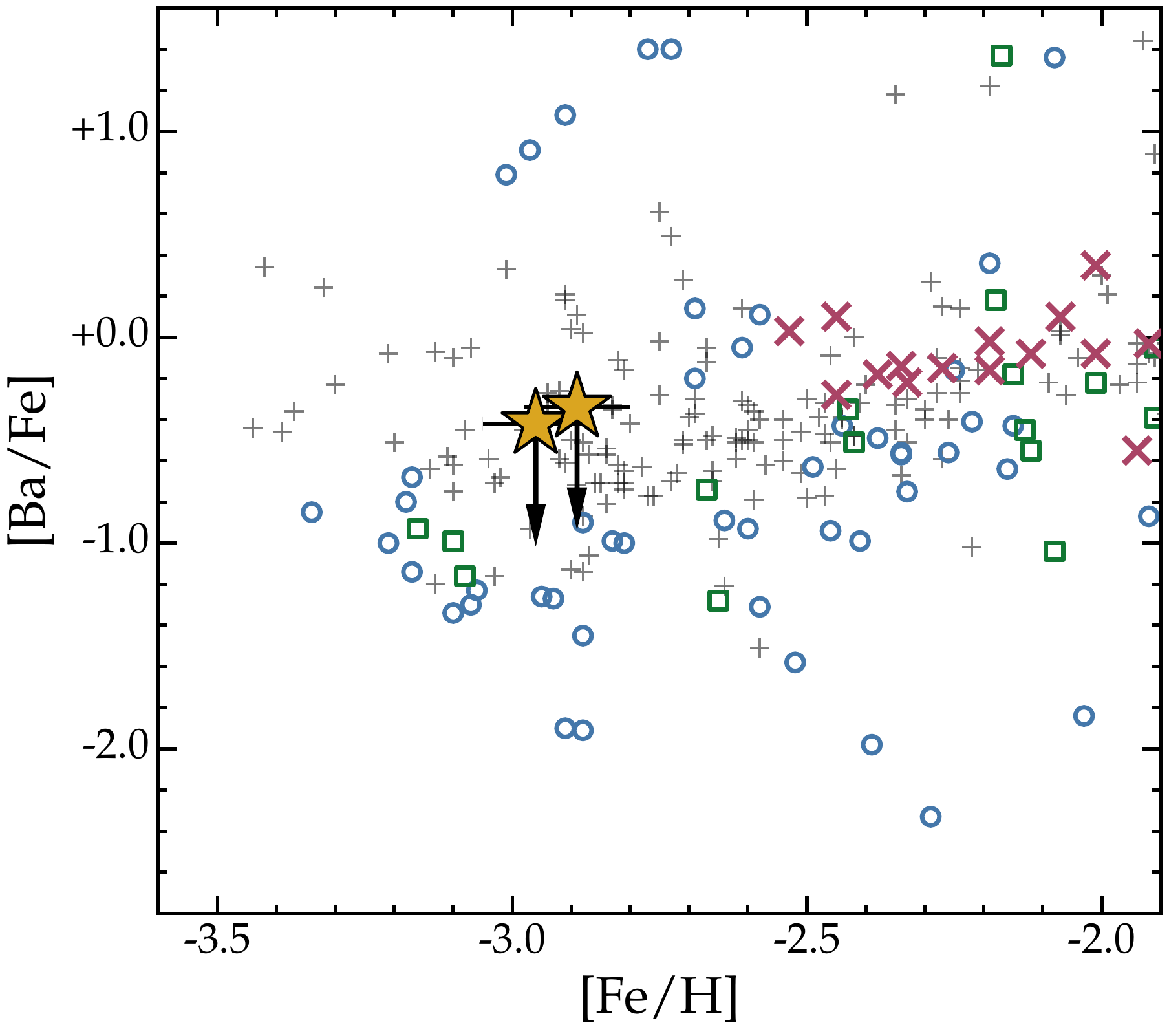} 
\end{center}
\vspace{-4mm}
\caption{
\label{alphaplot}
Comparison of abundances in Sylgr stream stars (gold stars) 
with abundances in stars in other stellar populations.
Abundances in stars in the 
\object[NAME UMi Galaxy]{UMi} dSph galaxy 
are shown by green squares.
Abundances in stars in 15~UFD galaxies
are shown by blue circles.
The mean abundance ratios within 18~GCs
are shown by red crosses.
The field stars,
which include dwarf and subgiant stars
with \teff~$>$~5600~K and \logg~$>$3.6,
are shown by small gray crosses.
Duplicate results have been removed.
References for the comparison samples are given in
Appendix~\ref{litappendix}.~
Note that the vertical axis on the [Sr/Fe] and [Ba/Fe] panels
spans 4.4~dex, twice that of
the other panels.
 }
\end{figure*}

We detect lines of 11~metals from sodium (Na, $Z =$~11) 
to nickel (Ni, $Z =$~28)
in the spectra of \jtwo\ and \jeight.
Figure~\ref{alphaplot} compares the [X/Fe] ratios for each element X
to the ratios found in metal-poor stars in 
ultra-faint dwarf (UFD) galaxies,
which are low-luminosity galaxies with $M_V > -$7;
the \object[NAME UMi Galaxy]{Ursa Minor} (UMi) dwarf spheroidal (dSph) galaxy,
which represents the low-luminosity ($M_V = -$8.8) 
and low-mass ($M_{*} = 2.9 \times 10^{5}$~\msun; \citealt{mcconnachie12}) 
end of the classical dwarf galaxies;
the mean abundance ratios found in metal-poor Galactic GCs;
and metal-poor halo stars in the Solar neighborhood.
Most of the comparison samples report abundances derived
assuming LTE, so the abundance ratios in Sylgr stream stars
presented in Figure~\ref{alphaplot} are the ones computed
in LTE, except for Na, Mg, and Fe.

The two Sylgr stream stars have statistically
identical abundances for all elements detected in our spectra.
Quantitatively, the 
$\log\epsilon$ abundances ([X/Fe] ratios)
differ by less than $1.2\sigma$ ($2\sigma$),
and no differences exceed 0.13~dex (0.20~dex).

These two stars show an enhancement of the $\alpha$ elements
magnesium (Mg, $Z =$~12), 
silicon (Si, $Z =$~14), and 
calcium (Ca, $Z =$~20) relative to Fe,
[$\alpha$/Fe]~$\approx +0.32 \pm 0.06$.
This ratio is typical among
stars with [Fe/H]~$= -$3 in dSph and UFD galaxies and the halo,
although there is considerable scatter 
within the dwarf galaxies.
The most metal-poor GCs 
also exhibit a similar level of $\alpha$ enhancement.

The scandium (Sc, $Z =$~21) and titanium (Ti, $Z =$~22) abundances
are within the range found in
the comparison samples, but Figure~\ref{alphaplot} shows that
they lie on the high side
of the [Sc/Fe] and [Ti/Fe] distributions.
\citet{sneden16} noticed that the abundances of 
Sc, Ti, and the neighboring element vanadium (V, $Z =$~23)
are often
correlated in metal-poor stars.
The super-Solar ratios in the
two Sylgr stream stars follow the same pattern, 
with average [Sc/Fe]~$= +0.30 \pm 0.10$ and 
[Ti/Fe]~$= +0.46 \pm 0.09$.
Our upper limits on the [V/Fe] ratios are,
unfortunately, uninformative
([V/Fe]~$< +$1.3).

Ratios among all other elements in the Fe group that we have detected
in these two stars
(chromium, Cr, $Z =$~24; 
manganese, Mn, $Z =$~25; and Ni)
are typical.
Figure~\ref{alphaplot} demonstrates that 
there are no distinguishing characteristics among the
abundances of these elements relative to the comparison samples.
Our upper limits on the cobalt (Co, $Z =$~27) and
zinc (Zn, $Z =$~30) abundances
are high and uninformative
([Co/Fe]~$< +$0.7; [Zn/Fe]~$< +$1.3).

\subsection{Strontium and Barium}
\label{srba}

We detect the heavy element strontium (Sr, $Z =$~38) 
in the spectra of both \jtwo\ and \jeight.
The average ratio,
$\mathrm{[Sr/Fe]}=+0.22 \pm 0.11$,
is mildly super-Solar.
We do not detect barium (Ba, $Z =$~56) in either star.
We place
meaningful upper limits on the Ba abundance ratios,
[Ba/Fe]~$< -$0.42 in \jtwo\ and [Ba/Fe]~$< -$0.34 in \jeight,
demonstrating that Ba is not enhanced in these stars.

\section{Discussion}
\label{discussion}

GCs and dwarf galaxies experience very different star formation histories
that affect the chemical compositions of their stars.
Most GCs are chemically homogeneous except for a few light elements.
Stars in dwarf galaxies, on the other hand, exhibit a wide range
of metallicities and ratios among their metals.
A quick scan of figures~2 and 4 of 
\citet{ji19gru1tri2}, for example, reveals that
it is rare---but not unprecedented---for 
two randomly-selected stars
within a given UFD galaxy to have the same composition.
We explore in this section
how the chemistry of the two Sylgr stream stars compares to
that of metal-poor field stars and
surviving GCs, UFD galaxies, and dSph galaxies.
These ratios set empirical constraints
on the yields of early supernovae 
in the progenitor of the Sylgr stream.
We also provide order-of-magnitude limits
on the density and mass of the progenitor system,
and we discuss the implications of our results.

\subsection{[Fe/H]}
\label{cluesmetallicity}

\jtwo\ was targeted
by the SEGUE-1 survey because its colors were
consistent with being an F turnoff star.
The number of such candidates available in a given SEGUE field
far exceeded the number of available fibers, 
and objects in this class were preferentially selected to be
bluer---which served as a proxy for low metallicity---and brighter
\citep{yanny09}.
\jeight\ has virtually identical colors, and it was selected
as a quality assurance star in SEGUE-2.~
We selected these stars for observation based on 
the availability of \rv\ measurements from SEGUE
to confirm their membership in the Sylgr stream.
If other, more metal-rich candidate members of the Sylgr stream had not 
been selected for SEGUE spectroscopy based on their 
redder colors, then our sample
could be biased toward lower-metallicity stars.
We thus exercise caution while using 
the metallicities of these two stars
to infer the metallicity distribution function (MDF)
as a characteristic of the nature of the progenitor system.

The mass-metallicity relation for dwarf galaxies in the Local Group
(e.g., \citealt{kirby13massmetal})
predicts that a progenitor system with a mean 
$\feh\approx -2.9$ should have had
an extremely low stellar mass, $M_{*} < 10^2$~\msun.
This mass is lower than the mass of the stars
already identified by \citetalias{ibata19a} 
as candidate members of the Sylgr stream. 
The actual progenitor mass is likely to be significantly larger
than this estimate. 
In fact, no known dwarf galaxy has mean $\feh$ below $-2.7$, 
which makes the extrapolated value of the mass very uncertain.

\begin{figure}
\includegraphics[width=\columnwidth]{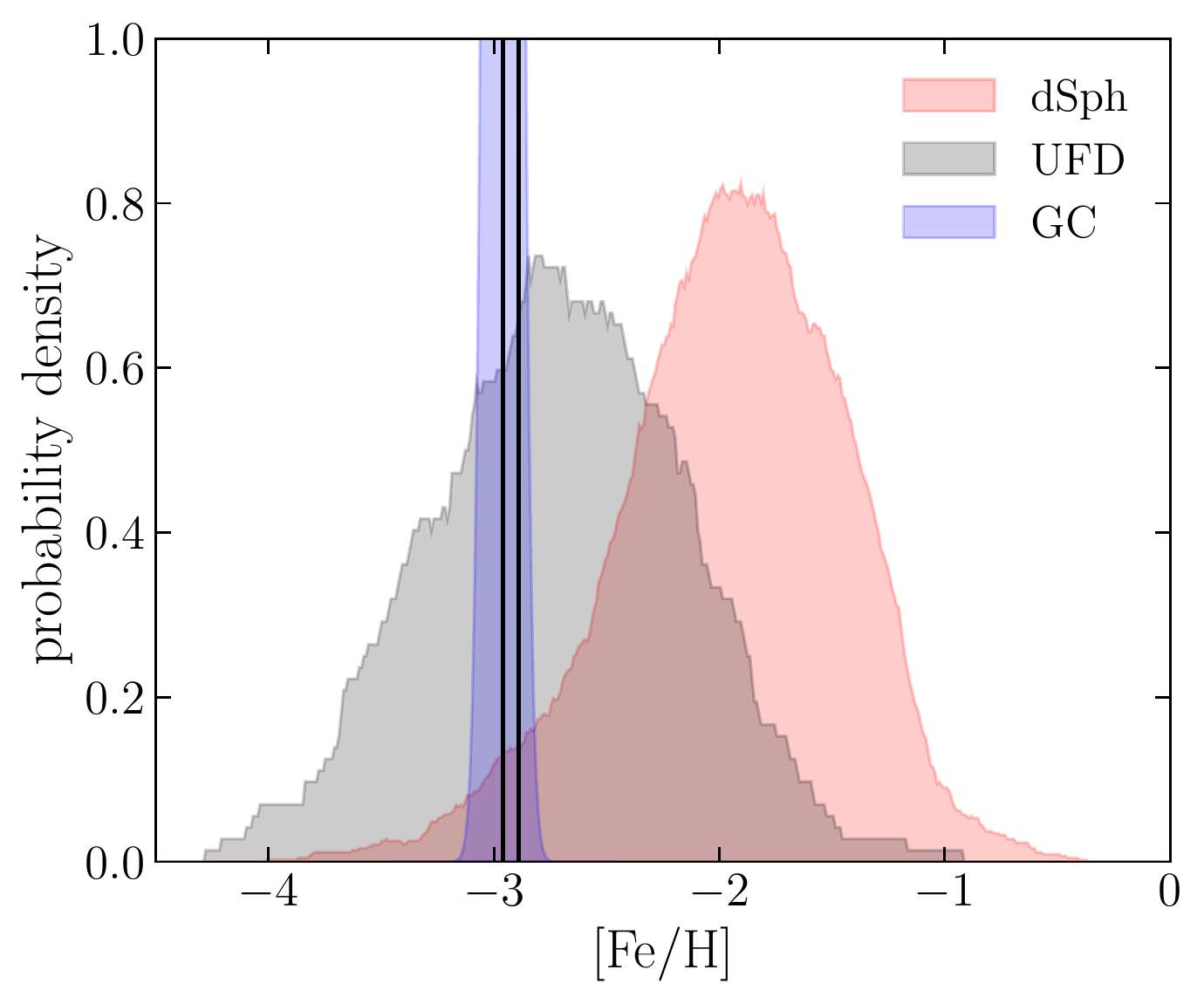}
\vspace{-6mm}
\caption{MDF for the sample of 1083~stars in the five smallest dSph galaxies from \citet{kirby10mdf} and the 72~stars from UFD galaxies described in Figure~\ref{alphaplot}. The inferred metallicities of the two Sylgr stars are shown by the vertical solid lines. The blue shaded histogram shows a Gaussian distribution, centered on \jtwo, with a standard deviation of 0.05~dex typical for GCs.}
\label{fig:mdf}
\end{figure}

While our sample may underestimate the mean metallicity 
of the progenitor system, as discussed above, 
we can still estimate the probability of having drawn 
these two stars from the MDF of a progenitor 
like one of the surviving dSph or UFD galaxies.
Figure~\ref{fig:mdf} shows the empirical MDFs 
for the five dSph galaxies with 
$M_{*} < 10^{6}$~\msun\
(\object[NAME CVn I dSph]{CVn~I},
\object[NAME Dra dSph]{Dra},
\object[NAME Leo II dSph]{Leo~II},
\object[NAME Sextans Dwarf Galaxy]{Sex}, and
\object[NAME UMi Galaxy]{UMi}) from \citet{kirby10mdf}.
The probability of randomly selecting two stars with 
$\feh=-2.96$ and $-2.89$ is below 0.16\%.
To approximate the MDF of UFD galaxies, 
we compile all 72~stars observed with high spectral resolution 
in the 15~UFD galaxies referenced in Appendix~\ref{litappendix}.
The probability of randomly selecting 
our two stars from this UFD sample is about 12\%.
Therefore, it is significantly more likely 
that the two stars could be chosen randomly 
from a UFD galaxy than from a classical dSph galaxy.

These probabilities do not include the more puzzling result that the inferred metallicities of the two stars are very close. We estimate $\Delta\feh \approx 0.07$~dex, based on the most accurate non-LTE abundance determination (Table~\ref{abundtab}). The probability of randomly selecting the first star from the UFD MDF and then selecting the second star with metallicity within $\Delta\feh$ is only 3.1\% (and much less for the dSph MDF).
This conditional probability is low because the MDFs of dwarf galaxies have significant intrinsic width (about 0.5~dex) and two random stars are unlikely to have such close [Fe/H] values.

In contrast, the MDF of GCs is very narrow, 
with typical standard deviation below 0.05~dex,
and often below observational errors
(e.g., \citealt{carretta09feh}). 
The two stars would naturally have similar values of $\feh$ 
if they belonged to a now-disrupted GC.~ 
However, such an origin of the Sylgr stream presents additional puzzles, which we discuss in Section~\ref{sec:implications} below.

\subsection{Lithium}
\label{clueslithium}

The Li abundance in the Sylgr stream stars,
$\log\epsilon = 2.05 \pm 0.07$,
is slightly lower than the mean Li abundances in the two 
GCs where Li abundances in dwarf stars have been studied,
\object[M30]{M30} 
($\log\epsilon =$~2.17, $\sigma = 0.16$~dex; \citealt{gruyters16})
and
\object[NGC 6397]{NGC~6397}
($\log\epsilon =$~2.24, $\sigma = 0.09$~dex; \citealt{lind09n6397}).
The probabilities of drawing the Li abundances in both 
Sylgr stream stars from either the 
\object[M30]{M30} or
\object[NGC 6397]{NGC~6397}
Li distributions are $\approx$~4.3\% and 0.10\%, respectively.
These results are sensitive to the 
methods of deriving \teff;
\citet{charbonnel05}, for example, note that 
$\log\epsilon$(Li) will change by $\pm 0.05$~dex
in response to changes of $\pm 70$~K in \teff.
\teff\ zeropoints can easily vary by $\pm 100$~K 
from one scale to another
(cf.\ Section~\ref{temperature}), 
and a reduction in the GC stars' 
\teff\ by 100~K would increase the 
probabilities to 10\% for
\object[M30]{M30} and 0.70\% for 
\object[NGC 6397]{NGC~6397}.
We conclude that the Li abundances of the Sylgr stream stars
are not significantly different from those of the GC stars.

No Li detections have been made in any 
dwarf stars in dSph or UFD galaxies,
because these systems are located at large distances
and their dwarf stars are faint.
Li abundances 
($\log\epsilon =$~2.19, $\sigma = 0.14$~dex; \citealt{monaco10})
in dwarf stars in the metal-poor
([Fe/H]~$\approx -1.75$) populations of the GC 
\object[NGC 5139]{$\omega$~Cen}, which may be the 
nuclear remnant of a dwarf galaxy
(e.g., \citealt{bekki03,ibata19omegacen}), 
are also similar to the GC and field star populations.

Li has been studied previously in dwarf stars in only
one metal-poor stellar stream.
\citet{roederer10a} presented Li abundances for
two dwarf stars that are probable 
members of the stream discovered by \citet{helmi99},
which was likely formed by a dwarf galaxy
with stellar mass $\sim 10^{8}$~\msun\
that was accreted by the Milky Way $\sim$~5--8 Gyr ago
\citep{koppelman19}.
These results are shown in Figure~\ref{lifeplot},
and they are also consistent with the Li abundances 
in other field stars.

Environment may not be a dominant factor in determining 
the Li content of metal-poor stars.
The \object[NGC 5139]{$\omega$~Cen},
Helmi stream, and Sylgr stream results
suggest that the primordial Li abundances in the progenitor 
systems were not remarkably different from those in the progenitors of
other systems whose stars were dispersed to form the nearby stellar halo.
Furthermore,
\citet{boesgaard05} and \citet{aoki09li}
found no significant correlation between orbital properties
and Li abundances in metal-poor dwarfs.
It will be important in future work
to assess the distribution
of Li abundances in larger samples of dwarf stars in Sylgr and 
other nearby stellar streams.
These streams both serve
as proxies for more distant, surviving systems, 
and they offer a new opportunity to study
the role of environment in Li production and depletion
in the early Universe.

\subsection{$\alpha$ Elements and Fe-Group Elements}
\label{cluesalphafe}

The Sylgr stream stars are $\alpha$ enhanced, with an average
enhancement of
[$\alpha$/Fe]~$\approx +0.32 \pm 0.06$
among Mg, Si, and Ca.
This result is typical for GCs and
the most metal-poor stars in 
UFD and dSph galaxies.
The ratios among Fe-group elements
(Sc, Ti, Cr, Mn, and Ni) are also typical 
for stars in UFD and dSph galaxies and GCs.~

The fact that both stars share ratios typical for other populations
implies that ejecta 
from multiple progenitor supernovae are present.
These abundance ratios reflect the mass-averaged supernova yields, 
rather than more extreme ratios that
result from stochastically sampling the high-mass end of the
initial mass function.
The enhanced [$\alpha$/Fe] ratios indicate 
that the stars formed
from gas enriched by the ejecta of 
Type~II supernovae, with minimal contributions
from Type~Ia supernovae,
so they probably formed shortly 
after star formation commenced.
The abundance ratios
do not directly reveal the stellar ages,
but they suggest that the stars
likely formed in an environment with a relatively
high star formation rate and short chemical enrichment timescale.

\subsection{Neutron-Capture Elements}
\label{cluesncap}

The Sr abundance
([Sr/Fe]~$= +0.22 \pm 0.11$)
and Ba upper limit 
([Ba/Fe]~$< -0.4$)
in \jtwo\ and \jeight\
indicate that these two stars are not highly enhanced
in elements produced by neutron-capture reactions.
The low Ba indicates that
neither star contains metals from a
low- or intermediate-mass ($\lesssim$~5~\msun)
companion star that passed through the
asymptotic giant branch phase of evolution.
Most stars with [Fe/H]~$\approx -3$
do not show evidence of such enrichment
(e.g., \citealt{jacobson15smss}),
so the stars are typical in this regard.
Otherwise, with such limited information 
it is not possible to determine 
what kind of nucleosynthesis process
produced the Sr in these stars.

Figure~\ref{alphaplot} shows that the [Sr/Fe] ratios
in the Sylgr stream stars
lie in the high part of the
distribution of stars in UFD galaxies.
Most [Sr/Fe] ratios in UFD galaxies are $< -$1.
Two notable exceptions include
\object[NAME Reticulum II]{Ret~II}
([Sr/Fe]~$= +$0.24; \citealt{ji16ret2}),
and 
\object[NAME Tucana III]{Tuc~III}
([Sr/Fe]~$=-$0.11; 
\citealt{hansen17tuc3,marshall19}),
which are known to be \rpro-enhanced galaxies.
These two galaxies are also enhanced to varying degrees
in other heavy elements produced by the \rpro, like Ba
([Ba/Fe]~$= +$1.04 and
$-$0.03, respectively).
If the Sylgr stream stars are also \rpro\ enhanced, 
their level of enhancement is considerably lower than either of 
\object[NAME Reticulum II]{Ret~II} or
\object[NAME Tucana III]{Tuc~III}.
We do not detect the \rpro\ element europium (Eu, $Z =$~63) 
in either spectrum,
and the upper limits we derive ([Eu/Fe]~$< +$2.04)
are uninformative.
The only other Sr-enhanced star known in a UFD galaxy
is in 
\object[NAME CVn II dSph]{CVn~II}
([Sr/Fe]~$= +$1.32; \citealt{francois16}),
which has one of the highest [Sr/Ba] ratios known
($+$2.6; \citealt{francois16}).
[Sr/Fe] ratios lower than the Sylgr stream stars
are found in 11/12 (92\%) 
of the non-\rpro-enhanced 
UFD galaxies studied at present.

Stars with [Fe/H]~$\sim -$3 in some dSph galaxies like 
\object[NAME Sextans Dwarf Galaxy]{Sextans} and
\object[NAME UMi Galaxy]{Ursa Minor}
show [Sr/Fe]~$\sim$~0 and 
[Ba/Fe]~$\sim -$1
\citep{cohen10umi,tafelmeyer10,kirby12},
potentially similar to the two stars observed in the Sylgr stream.
In other dSph galaxies, like
\object[NAME Carina dSph]{Carina} or
\object[NAME Dra dSph]{Draco},
[Sr/Fe] and [Ba/Fe] are both significantly sub-Solar
\citep{cohen09dra,venn12},
sometimes by several dex
\citep{fulbright04dra}.

Sr is not often studied in GCs.
Figure~\ref{alphaplot} shows that 5/7 (71\%) of the
metal-poor GCs ([Fe/H]~$< -1.8$) 
show mean [Sr/Fe] ratios $< 0$, 
lower than the Sylgr stream stars.
Ba is studied more frequently than Sr in GCs,
and 14/16 (88\%) of the GCs
have [Ba/Fe] ratios
higher than the [Ba/Fe] upper limits
we have derived.
Thus the enhanced [Sr/Ba] ratio
we derive for the Sylgr stream, $> +0.6$,
is unprecedented in metal-poor GCs.

In conclusion,
the [Sr/Fe], [Ba/Fe], and [Sr/Ba] ratios 
found in the Sylgr stream stars are rare 
among stars in UFD galaxies, sometimes found
among the most metal-poor stars
in dSph galaxies, and not found in metal-poor GCs.~

\subsection{Light Elements Whose Abundances Vary in GCs}
\label{cluesnaal}

Particular abundance variations among some light elements
are a sign of GC populations.
Star-to-star scatter exists
among the O, Na, and aluminum (Al, $Z =$~13) abundances in stars
in GCs, with the lowest [Na/Fe] and [Al/Fe] ratios
coincident with those in field stars and the
highest [Na/Fe] and [Al/Fe] ratios enhanced by a dex or more.
Neither ratio is significantly enhanced in \jtwo\ or \jeight.
The typical [O/Fe] ratios found in metal-poor GC stars 
are $\leq +0.8$ or so 
(e.g., \citealt{carretta09ona}),
considerably lower than the upper limits we derive 
([O/Fe]~$< +1.5$).
The [Mg/Fe] ratios in some stars in a few GCs
are depleted,
but the [Mg/Fe] ratios in \jtwo\ and \jeight\
do not show evidence of such depletion.
These ratios are all consistent with those
found in typical metal-poor field stars
and so-called ``first-generation'' stars in GCs, 
and they do not exclude the possibility 
that the progenitor was a GC.

\subsection{Other Constraints on the Nature of the Progenitor}
\label{sec:density}

Many GCs and UFD galaxies have 
velocity dispersions $\sim 5$~\kmsec\ 
(e.g., \citealt{harris96,simon07}).
The small \rv\ difference between
\jtwo\ and \jeight\ ($6.5\pm0.5$~\kmsec)
cannot distinguish between GC and UFD galaxy populations,
especially in an extended stellar stream whose members
are not in dynamical equilibrium.

\citetalias{ibata19a} note that the Sylgr stream has similar 
energy and angular momentum to the GCs
\object[M10]{M10} (NGC~6254) and 
\object[M12]{M12} (NGC~6218). 
Neither cluster has an obvious excess of 
extra-tidal stars \citep{deboer_etal19,kundu19}.
\citetalias{ibata19a} excluded these GCs
as the progenitor based on their 
less radial orbits with smaller apocenters 
\citep[$< 5$~kpc;][]{baumgardt_etal19}
and
the much higher metallicities of
\object[M10]{M10} and
\object[M12]{M12}
($\feh \approx -1.6$ and $-1.3$; \citealt{carretta09feh})
relative to the SSPP metallicity estimates.
We reaffirm this conclusion on the basis of
the low metallicities 
derived from our high-resolution spectra.

\citetalias{ibata19a} identified the Sylgr stream as a long arc,
suggesting that the disruption 
took place no more than a few orbital periods ago.
The orbit reconstruction by \citetalias{ibata19a} 
suggests a bound orbit for the Sylgr stream, 
with pericenter near 2.5~kpc from the Galactic center 
and apocenter near 20~kpc.
Such an orbit has a relatively short period 
$\la 0.5$~Gyr and can be reliably calculated, 
as the gravitational potential of the Galaxy 
is likely to be stable over many orbital periods.
This is a typical prograde orbit shared by many surviving GCs.~ 
In contrast, the pericenter of this orbit 
is much smaller than typical pericenters for UFD galaxies 
(20--40~kpc; \citealt{fritz18, simon18}). 
Therefore, the progenitor experienced significantly stronger tidal forces 
than the current population of 
UFD galaxies, 
so it would have needed to be more dense than the UFD galaxies
to avoid an early disruption.

The average density of Galactic material at $r=2.5$~kpc 
from the center is $\sim 1$~\msun~pc$^{-3}$. 
An object would remain gravitationally bound for several orbital periods 
if its average density exceeds roughly twice this value. 
GCs with orbital pericenters $\approx$2--3~kpc \citep{helmi18}
have stellar densities at their half-mass radii
$\approx$~10--500~\msun~pc$^{-3}$, 
with a median value $\approx$~100~\msun~pc$^{-3}$
\citep{baumgardt_hilker18},
so they easily meet this condition.
Surviving UFD galaxies have average densities
$\ll 1$~\msun~pc$^{-3}$ within their core radii \citep{simon07},
so any UFD-like systems on orbits like that of the Sylgr stream
would have disrupted long ago
(cf.\ \citealt{roederer18d}).
We conclude that the progenitor must have been substantially 
more dense than the surviving population of UFD galaxies.

We can estimate the minimum mass of the progenitor 
required to withstand the tidal forces along its orbit. 
If the Sylgr progenitor had a half-mass radius of $\approx$~6~pc,
similar to the surviving GCs with comparable orbital pericenters,
then it would need to have had a minimum mass of 
$\sim 10^3$~\msun\ before disruption.
The initial mass of the cluster on such an orbit 
was likely up to a factor of 10 higher \citep{baumgardt_etal19}. 
This order-of-magnitude estimate shows that the progenitor 
could contain enough stars 
to explain all stream candidates 
and persist for many orbital periods.

\subsection{Implications If the Progenitor Was a GC}
\label{sec:implications}

If the progenitor of the Sylgr stream was indeed a GC, 
it would be an exciting discovery. 
The [Fe/H] ratios of the two observed stars are
considerably lower than those of
the lowest-metallicity Galactic GCs,
which have [Fe/H]~$\approx -$2.4 \citep[e.g.,][]{harris96}.
\citet{beasley19} considered GC systems of external galaxies 
that span 6 orders of magnitude in stellar mass and
concluded that the MDFs of these systems have a plausible 
``metallicity floor'' at $\feh\approx -2.5$.
Few known GCs show metallicity below that value \citep{forbes_etal18}.
The two stars observed in the Sylgr stream have Fe abundances that are
a factor of $\approx 2.5$ lower than this floor.
Although the [Fe/H] scales may systematically differ by about 0.1~dex 
\citep[e.g.,][]{carretta09feh},
the metallicity of the two Sylgr stream stars
is demonstrably lower than that of all known GCs.~

The mass-metallicity relation for galaxies is observed to hold with only mild evolution even at high redshift \citep[e.g.,][]{maiolino_mannucci19}. Thus the presence of a metallicity floor indicates that surviving GCs could not have formed in galaxies below a certain mass. That minimum mass is estimated to be around halo mass $M_{\rm h} \sim 10^9$\msun\ \citep{choksi_etal18} or stellar mass $M_* \sim 10^6$\msun\ \citep{kruijssen19}. The exact values are very sensitive to the unknown extrapolation of the galactic mass-metallicity relation at $z>4$, when the most metal-poor GCs are expected to form \citep[e.g.,][]{muratov_gnedin10, li_gnedin14, choksi_gnedin19}.

Confirming the origin of the Sylgr stream progenitor as a GC would have significant implications for the universality of such a floor. It would indicate that GCs can form in lower-mass galaxies than had been expected previously, which is an important constraint for models of GC formation. It would also imply that a non-zero fraction of halo stars with [Fe/H]~$< -$2.5 formed in GCs, rather than in situ or in UFD or dSph galaxies.

Even if the Sylgr progenitor was instead a UFD or dSph galaxy,
the arguments in Section~\ref{sec:density} indicate that 
the observed thin stream is likely a remnant of its densest part. 
This dense part could be a nuclear star cluster, 
with a stellar density similar to that of massive GCs.

\section{Conclusions}
\label{conclusions}

We have obtained high-resolution optical spectra of
two warm, main sequence stars 
proposed by \citetalias{ibata19a} as members of the
Sylgr stellar stream.
We confirm previous \rv\ measurements,
verifying their status as members.
We have derived abundances of 13~elements in each star,
and we show that these two stars are chemically homogeneous
at a level of $\leq 0.13$~dex.
Both stars are metal-poor, with average
[Fe/H]~$= -2.92 \pm 0.06$.
The Li abundances,
$\log\epsilon$(Li)~$= 2.05 \pm 0.07$,
are consistent with other unevolved
stars at this metallicity.
Neither star is C enhanced, with
[C/Fe]~$< +1.0$.
Both stars are $\alpha$ enhanced, with average
[$\alpha$/Fe]~$\approx +0.32 \pm 0.06$
among the $\alpha$ elements Mg, Si, and Ca.
The ratios among other elements
(Na, Al, Sc, Ti, Cr, Mn, and Ni) are typical
for stars at this metallicity.
Sr is mildly enhanced, with average
[Sr/Fe]~$= +0.22 \pm 0.11$,
but Ba is not, with 
[Ba/Fe]~$< -0.4$.

We compare the chemical composition of these two stars with
that of other metal-poor stars in the field and in
surviving stellar systems around the Milky Way.
Our results indicate that 
the progenitor of the Sylgr stream could have been
similar to a GC or 
a low-mass UFD or dSph galaxy.
In either case, however, some abundances
are unlike those found in surviving GCs and UFD and dSph galaxies.
For example, the high [Sr/Ba] ratio 
is not found in GC populations.
On the other hand,
the fact that both stars have 
identical metal abundances
favors a GC origin.
If the progenitor was a GC, it would qualify as the 
most metal-poor GC known, by about 0.4~dex.
Dynamical considerations from the orbit of the Sylgr stream
also favor a system with a relatively high stellar density,
like a GC or a dense region of a UFD or dSph galaxy.

With only two stars in our sample, these considerations remain unconfirmed. Spectroscopic observations of more stars in the Sylgr stream are needed to conclusively determine its origin. If more stars are found to have nearly the same metallicity, it would very likely confirm a GC origin. Similarly, if other stream stars have different metallicities but a substantial fraction show indistinguishable values of $\feh \approx -3$, then the latter subset of stars would still point to a disrupted GC or nuclear star cluster. Such a discovery would challenge the apparent metallicity floor of GCs and present an exciting puzzle for the theories of GC formation.

\acknowledgments

I.U.R.\ acknowledges financial support from
grants AST~16-13536, AST-1815403,
and PHY~14-30152 (Physics Frontier Center/JINA-CEE)
awarded by the U.S.\ National Science Foundation (NSF).
O.Y.G.\ was supported in part by the NSF through grant AST~14-12144.~
We thank the referee for a helpful report.
We also thank
E.~Bell, T.S.\ Li, K.~Malhan, M.~Mateo, and M.~Valluri
for helpful discussions, and
C.\ Sneden and V.\ Placco for
developing and maintaining the LINEMAKE code
(\url{https://github.com/vmplacco/linemake}).
This research has made use of NASA's
Astrophysics Data System Bibliographic Services;
the arXiv pre-print server operated by Cornell University;
the SIMBAD and VizieR databases hosted by the
Strasbourg Astronomical Data Center;
the Atomic Spectra Database hosted by 
the National Institute of Standards and Technology;
the INSPECT database
(v.\ 1.0, \url{http://www.inspect-stars.com});
and
the Image Reduction and Analysis Facility (IRAF) software packages
distributed by the National Optical Astronomy Observatories,
which are operated by AURA,
under cooperative agreement with the NSF.~
This work has also made use of data from the European Space Agency (ESA)
mission {\it Gaia} 
(\url{http://www.cosmos.esa.int/gaia}), 
processed by the {\it Gaia} Data Processing and Analysis Consortium (DPAC,
\url{http://www.cosmos.esa.int/web/gaia/dpac/consortium}). 
Funding for the DPAC has been provided by national institutions, in particular
the institutions participating in the {\it Gaia} Multilateral Agreement.

\facility{Magellan (MIKE)}

\software{IRAF \citep{tody93},
matplotlib \citep{hunter07},
MOOG \citep{sneden73},
numpy \citep{vanderwalt11},
R \citep{rsoftware},
scipy \citep{jones01}}

\appendix
\section{References for Literature Data}
\label{litappendix}

The abundance data for the comparison samples of stars
presented in Figure~\ref{alphaplot} have been
compiled from many sources.  
Abundances in stars in the 
\object[NAME UMi Galaxy]{UMi} dSph galaxy 
are taken from
\citet{shetrone01},
\citet{sadakane04},
\citet{cohen10umi},
\citet{kirby12}, and 
\citet{ural15}.
Abundances in stars in the UFD galaxies 
\object[NAME Bootes Dwarf Spheroidal Galaxy]{Boo~I},
\object[NAME Bootes II]{Boo~II},
\object[NAME Coma Dwarf Galaxy]{Com},               
\object[NAME CVn II dSph]{CVn~II},
\object[NAME Grus I]{Gru~I},
\object[NAME Her Dwarf Galaxy]{Her},                
\object[NAME Hor I]{Hor~I},                         
\object[NAME Leo IV Dwarf Galaxy]{Leo~IV},          
\object[NAME Reticulum II]{Ret~II},                 
\object[NAME Segue 1]{Seg~1},                       
\object[NAME Segue 2]{Seg~2},     
\object[NAME Tri II]{Tri~II},                  
\object[NAME Tucana II]{Tuc~II},                    
\object[NAME Tucana III]{Tuc~III},                  
and
\object[NAME UMa II Galaxy]{UMa~II}
are taken from
\citet{koch08her,koch13her},
\citet{feltzing09},
\citet{frebel10ufd,frebel14seg1,frebel16boo1},
\citet{norris10seg1,norris10boo},
\citet{simon10},
\citet{gilmore13},
\citet{ishigaki14},
\citet{koch14boo2},
\citet{roederer14b},
\citet{francois16},
\citet{ji16tuc2,ji16ret2,ji16boo2,ji19gru1tri2},
\citet{roederer16b},
\citet{hansen17tuc3},
\citet{chiti18},
\citet{nagasawa18}, and 
\citet{marshall19}.
The mean abundance ratios within GCs 
including
\object[M15]{M15},
\object[M30]{M30},
\object[M53]{M53},
\object[M55]{M55},
\object[M68]{M68},
\object[M92]{M92},
\object[NGC 2419]{NGC~2419},
\object[NGC 4372]{NGC~4372},
\object[NGC 4833]{NGC~4833},
\object[NGC 5053]{NGC~5053},
\object[NGC 5694]{NGC~5694},
\object[NGC 5824]{NGC~5824},
\object[NGC 5897]{NGC~5897},
\object[NGC 6287]{NGC~6287},
\object[NGC 6397]{NGC~6397},
\object[NGC 6426]{NGC~6426},
\object[NGC 6535]{NGC~6535},
and
\object[Cl Terzan 8]{Ter~8}, 
are computed from data presented in
\citet{shetrone01},
\citet{lee02},
\citet{cohen11m92},
\citet{cohen11n2419},
\citet{koch11,koch14n5897},
\citet{sobeck11},
\citet{venn12},
\citet{mucciarelli13},
\citet{carretta14ter8},
\citet{boberg15,boberg16},
\citet{roederer15},
\citet{sanroman15},
\citet{schaeuble15}, 
\citet{roederer16a},
\citet{bragaglia17},
\citet{hanke17}, and
\citet{rain19}.
The field star sample 
(small gray crosses) includes 
stars with \teff~$>$~5600~K and \logg~$>$3.6 from 
\citet{lai08},
\citet{bonifacio09},
\citet{yong13a}, and
the main sequence and subgiant
stars analyzed by \citet{roederer14c}.
Duplicate results have been removed.

\bibliographystyle{aasjournal}
\bibliography{iuroederer,gc}

\end{document}